\documentclass[twocolumn,superscriptaddress,showpacs,preprintnumbers,amsmath,amssymb]{revtex4}
\usepackage{bm,graphicx,mhchem}
\newcommand{\bb}{\begin{equation}}
\newcommand{\en}{\end{equation}}
\newcommand{\Wr}{\overrightarrow{\omega}}
\newcommand{\Wl}{\overleftarrow{\omega}}
\newcommand{\Ur}{\overrightarrow{\rho}}
\newcommand{\Ul}{\overleftarrow{\rho}}

\begin{document}

\title{Fluctuation theorem and large deviation function \\
for a solvable model of a molecular motor}

\author{D. Lacoste}
\affiliation{Laboratoire de Physico-Chimie Th\'eorique, UMR 7083,
ESPCI, 10 rue Vauquelin, 75231 Paris cedex 05, France}
\author{A.W.C. Lau}
\affiliation{Department of Physics, Florida Atlantic University,
777 Glade Rd., Boca Raton, FL 33431}
\author{K. Mallick}
\affiliation{Institut de Physique Th\'eorique, CEA Saclay, 91191
Gif, France}

\begin{abstract}
We study a discrete stochastic model of a molecular motor. This
discrete model can be viewed as a \emph{minimal} ratchet model. We
extend our previous work on this model, by further investigating
the constraints imposed by the Fluctuation Theorem on the
operation of a molecular motor far from equilibrium. In this work,
we show the connections between different formulations of the
Fluctuation Theorem. One formulation concerns the generating
function of the currents while another one concerns the
corresponding large deviation function, which we have calculated
exactly for this model. A third formulation concerns the ratio of
the probability of observing a velocity $v$ to the same
probability of observing a velocity $-v$. Finally, we show that
all the formulations of the Fluctuation Theorem can be understood
from the notion of entropy production.
\end{abstract}

\date{\today}
\maketitle


\section{introduction}
The living cell has evolved a diverse array of proteins, which can
perform a variety of chemical functions. These proteins catalyze
chemical reactions, control key processes like recognition or
signalling, or act as molecular machines. Biological systems
containing these elements are typically described as active
because they are in a non-equilibrium state as opposed to
non-active systems which can be considered at thermal equilibrium.
Recently there has been a lot of interest in the non-equilibrium
properties of active biological systems, such as the hair bundle
of the ear \cite{martin,duke}, active membranes \cite{jb_PRE},
active gels \cite{joanny}, active networks \cite{fred}, active
lipid clusters \cite{madan_mayor} and living cells \cite{andy}...

Clearly, one step towards understanding active biological systems,
starts with the understanding of single active proteins, such as
for instance molecular motor proteins. These molecular motor are
nano-machines that convert chemical energy derived from the
hydrolysis of ATP into mechanical work and motion
\cite{howard,motorreview}. Important examples include kinesin,
myosin, RNA polymerase, and the $F_0/F_1$ rotating motor.
Mechano-transduction ({\it i.e.}, the process of conversion of
chemical energy into mechanical work) in motors has been described
theoretically by ratchet-models \cite{armand1,parmeggiani}, which
rely on the fruitful concept of broken symmetry. According to the
Curie principle, directed motion requires to break the spatial
symmetry and the time reversal symmetry associated with
equilibrium (detailed balance). The spatial symmetry is broken by
the asymmetric interaction between the motor and the filament,
while the time reversal symmetry is broken by chemical
transitions, which break locally the detailed balance condition.
From the continuous ratchet models described by Langevin equations
\cite{armand1}, it is possible to construct discrete stochastic
models of molecular motors by considering only localized discrete
transitions as explained in Ref.~\cite{widom}. These discrete
stochastic models are interesting because they are \emph{minimal},
in the sense that they contain the main physical picture of
ratchet models while being more amenable to precise mathematical
analysis \cite{kolomeisky,nelson,mazonka,lipowsky}.

Recent advances in experimental techniques
 to probe the fluctuations of single
motors provide ways to gain insight into their kinetic pathways
\cite{block,coppin,carter,schnitzer}. However, a general
description for fluctuations of systems driven out of equilibrium,
and in particular of motors, is still lacking. Recently, the
Fluctuation Theorem (FT)
 has emerged as a
promising framework to characterize fluctuations in
far-from-equilibrium regimes. This theorem is in fact a group of
closely related results valid for a large class of non-equilibrium
systems \cite{FT,gallavotti,lebowitz,evans,crooks}. In a nutshell,
FT states that the probability distribution for the entropy
production rate obeys a symmetry relation. The Theorem becomes
particulary relevant for small systems in which fluctuations are
large. For this reason, the FT has been verified in a number of
beautiful experiments with small systems such as biopolymers and
colloidal systems \cite{ritort}. In the specific case of molecular
motors, FT leads to constraints on the operation of molecular
motors or nano-machines far from equilibrium in a regime where the
usual thermodynamic laws do not apply \cite{qian,gaspard,seifert}.

In a recent work \cite{prl}, we have extended a two-states
discrete stochastic model introduced in Ref.\ \cite{nelson} by
including an important variable, namely the number of ATP
consumed. We have shown that this extended model satisfies FT, and
we have constructed a thermodynamic framework allowing us to
characterize quantities like the average velocity, the average ATP
consumption rate and its thermodynamic efficiency. We have also
analyzed the different thermodynamic modes of operation of the
motor as functions of generalized forces arbitrarily far from
equilibrium. Using FT, we have quantified the "violations" of
Einstein and Onsager relations. The deviations from Einstein and
Onsager relations can be studied by considering the linear
response theory in the vicinity a non-equilibrium steady state
rather than near an equilibrium steady state. After determining
the parameters of our model by a fit of single molecule
experiments with kinesin \cite{block}, we have formulated a number
of theoretical predictions for the "violations" of Einstein and
Onsager relations for this motor.

In this paper, we further extend the analysis of this model. In
particular, we provide a more detailed study of the modes of
operation of the motor and its thermodynamic efficiency in
relation with the experimental data of kinesin. This part contains
important information  which could not be presented in
Ref.~\cite{prl} due to limited space. The rest of the paper is
devoted to bringing together different formulations of FT,
explaining their connections and their physical implications. One
way to formulate FT involves the generating function of the
currents. We show that this formulation leads to two versions of
FT, a long time version of FT similar to the Gallavotti-Cohen
relation which holds quite generally, and a finite time version
which is analogous to the Crooks-Evans transient Fluctuation
Theorem \cite{crooks,evans} which holds under restricted
hypotheses for the initial state. Another formulation of FT takes
the form of a property of the large deviation function of the
current. There are very few non-equilibrium models for which the
large deviation function of the current is known exactly. Our
model is sufficiently simple for this analytical calculation to be
possible, and by carrying it out we show that the large deviation
function of the current satisfies indeed an FT relation. A
by-product of this calculation is a third formulation of FT in
terms of the ratio of probabilities for observing a velocity $v$
to the same probability for observing a velocity $-v$. The
prediction of this ratio of probabilities, which is obtained from
the large deviation function of the current is one of the main
results of this paper. We also study the connections between FT
and the notion of entropy production. This entropy production can
be explicitly calculated using the notion of affinities associated
with a cyclic evolution of the mechanical and chemical variables,
and the affinities precisely enter all the formulations of FT. We
show that the entropy production can also be obtained from an
evaluation of a quantity called the action functional
\cite{lebowitz}. The entropy production and therefore also FT
depends on the coarse-graining of the description which we
illustrate by considering three different levels of description:
purely mechanical, purely chemical and a combination of the two.

The paper is organized as follows: In Section II, we introduce the
model, in Section III we consider the modes of operation of the
motor, its thermodynamic efficiency and the comparison of the
model with experimental data for kinesin. Section IV is devoted to
the formulation of FT in terms of generating functions, with its
long time and its finite time versions, Section V discusses the
formulation in terms of the large deviation function of the
current and the significance of the third formulation of FT in
terms of a ratio of probabilities, and the last section contains
the discussion of the entropy production.

\section{A two-state model for molecular motors}
\subsection{Construction of the model}
\begin{figure}
\includegraphics[height=1.45in]{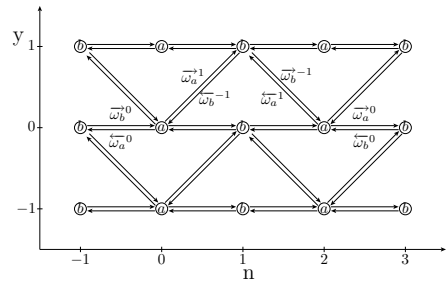}
\caption{A schematic of the rates for this two-state stochastic
model of a single processive molecular motor. The position of the
motor is $n$ and $y$ is the number of ATP consumed. The even and
odd sites are denoted by $a$ and $b$, respectively. In the case of
two headed kinesin, site $a$ represents a state where both heads
are bound to the filament, whereas site $b$ represents a state
with only one head bound. Note that the lattice of a and b sites
extend indefinitely in both directions along the $n$ and $y$ axis.
All the possible transitions are represented with arrows on this
particular section of the lattice.} \label{fig:sketch}
\end{figure}

\begin{figure}
\includegraphics[scale=0.8]{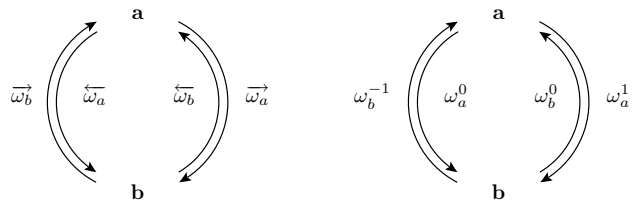}
\caption{Cycles associated with the evolution of the motor of
Fig.~\ref{fig:sketch}. Left: cycle for the position variable $n$;
the average length $n$ undergone by the motor corresponds to half
the number of full turns of the clock (the factor half is due to
the period  being  equal to two lattice units) and with the rates
as shown. Right: cycle for the chemical variable $y$, the average
number of ATP units corresponds to the number of full turns with
the rates shown. The affinities associated with these cycles are
given by  Eq.~\ref{Effective potential} for the mechanical
variable and  by Eq.~\ref{Effective potential_chimie} for the
chemical variable.} \label{fig:representation}
\end{figure}
As a result of conformational changes powered by hydrolysis of
ATP, a linear processive motor, like kinesin, moves along a
one-dimensional substrate (microtubule).  The state of the
molecular motor may be characterized by two variables: its
position and the number of ATP consumed. To model the dynamics, we
consider a linear discrete lattice, where the motor ``hops" from
one site to neighboring sites, either consuming or producing ATP
(see Fig.~\ref{fig:sketch}). An alternate representation of the
dynamics can be built in terms of cycles (see
Fig.~\ref{fig:representation}). The position is denoted by $x = n
d$, where $2 d \approx 8 \, \mbox{nm}$ is the step size for
kinesin. The even sites (denoted by $a$) are the low-energy state
of the motor, whereas the odd sites (denoted by $b$) are its
high-energy state; their energy difference is $ \Delta E \equiv
k_B T \epsilon$, where $k_B$ is the Boltzmann constant and $T$ is
the temperature. This model is suitable to describe a two-headed
kinesin walking on a microtubule, with the high-energy state
corresponding to the state where a single head is bound to the
filament, and the low-energy state corresponding to the two heads
bound to it. Because of the periodicity of the filament, all the
even ($a$) sites and all the odd ($b$) sites are equivalent.

The dynamics of the motor is governed by a master equation for the
probability, $P_i(n,y,t)$, that the motor, at time $t$, has
consumed $y$ units of ATP and is at site $i$ $(=a,b)$ with
position $n$:
\begin{eqnarray}
\partial_t P_i(n,y,t)  = -  \left( \overleftarrow{\omega}_i + \overrightarrow{\omega}_i
\right) P_{i}(n,y,t)+
\phantom{\overleftarrow{\omega}_{n+1}^{\,l}ppppppp}\nonumber \\
\sum_{l = -1,0,1} \left [\,
\overleftarrow{\omega}_{j}^{\,l}\,P_{j}(n+1,y-l,t)+
\overrightarrow{\omega}_{j}^{\,l}\,P_{j}(n-1,y-l,t) \,\right ],
\label{Master eq}
\end{eqnarray}
with $i \neq j$,   $\overleftarrow{\omega}_j \equiv \sum_l
\overleftarrow{\omega}_{j}^{\,l}$ and $\overrightarrow{\omega}_j
\equiv \sum_l \overrightarrow{\omega}_{j}^{\,l}$. We denote  by
$\overleftarrow{\omega}_{j}^{\,l}$ and
$\overrightarrow{\omega}_{j}^{\,l}$   the transition rates
  from a  site $j$ to a neighboring site to the left
or to the right, respectively, and  with $l\,(= -1,0,1)$ ATP
molecules consumed.   Note that the index $i$ is directly linked
with the parity of $n$. We expect the rates
$\overleftarrow{\omega}_{j}^{\,l}$ and
$\overrightarrow{\omega}_{j}^{\,l}$ to be different even if no
load is applied to the motor, because the interaction between the
motor and the filament breaks the left-right (spatial) symmetry.
This requirement is essential for generating directed motion.

Another  essential  requirement for generating directed motion is
to break  time-reversal symmetry. Such a symmetry is  always
present  at
 equilibrium, where  the detailed balance condition holds.
 Detailed balance is broken in molecular motors
due to chemical transitions involved in the mechano-transduction
process. To obtain a simple description of this process, we assume
that transitions between the two states $\mbox{M}_a$ and
$\mbox{M}_b$ of the motor \cite{parmeggiani} are possible via two
different chemical pathways:
\begin{equation}\label{chemical reaction 1}
\ce{M_a + ATP <=>[\alpha_a][\alpha_b] M_b + ADP + P},
\end{equation}
and
\begin{equation}\label{chemical reaction 2}
\ce{M_a  <=>[\beta_a][\beta_b] M_b},
\end{equation}
where $\alpha_a$ (resp. $\alpha_b$) are forward (resp. backward)
rates. The first pathway ($\alpha$) represents transitions of the
motor accompanied by ATP hydrolysis, which we call "active", and
the second pathway ($\beta$) represents transitions driven by
thermal activation, which we call "passive". In the representation
of fig.~\ref{fig:sketch}, the $\alpha$-pathway represents oblique
transitions which change both $y$ and $n$ whereas the
$\beta$-pathway is associated with horizontal transitions which
change only $n$. It is straightforward to generalize the model
with more chemical pathways, but here we focus only on these two
\cite{parmeggiani}. In the absence of an external force,
transition state theory of chemical reactions requires that
\cite{hill,deDonder} \bb
\frac{\alpha_a}{\alpha_b}=e^{-\epsilon+\Delta \mu},
\label{thermo_rates1} \en and  \bb
 \frac{\beta_a}{\beta_b}=e^{-\epsilon}. \label{thermo_rates2} \en
 Taking  $\alpha_a=\overrightarrow{\omega_a}^1,
\alpha_b=\overleftarrow{\omega_b}^{-1}, \beta_a=
\overrightarrow{\omega_a}^0, \beta_b=\overleftarrow{\omega_b}^0$
 and $\alpha_a=\overleftarrow{\omega_a}^1,
\alpha_b=\overrightarrow{\omega_b}^{-1}, \beta_a=
\overleftarrow{\omega_a}^0, \beta_b=\overrightarrow{\omega_b}^0$,
 we construct the transition  rates from
 only four unknown parameters $\alpha$, $\alpha'$, $\omega$ and
$\omega'$ as follows, \bb
\begin{array}{ll} \overleftarrow{\omega_b}^{-1} = \alpha\, , &
\overleftarrow{\omega_b}^0
 =  \omega\,,  \\
\overrightarrow{\omega_a}^1   =  \alpha\, e^{ -\epsilon + \Delta
\mu },
& \overrightarrow{\omega_a}^0 = \omega\,e^{-\epsilon },    \\
\overleftarrow{\omega_a}^1  =  \alpha' e^{ -\epsilon + \Delta
\mu},
 & \overleftarrow{\omega_a}^0  = \omega'\,e^{-\epsilon},\\
\overrightarrow{\omega_b}^{-1} = \alpha' , &
\overrightarrow{\omega_b}^0
  =  \omega'\,,
\end{array} \label{rates f=0} \en
and with
$\overleftarrow{\omega_b}^{1}=\overrightarrow{\omega_a}^{-1}
=\overleftarrow{\omega_a}^{-1}=\overrightarrow{\omega_b}^{1}=0$.
The only thermodynamic force driving the chemical cycle is the
free energy of hydrolysis. This is quantified by the chemical
potential $\Delta \widetilde{\mu} \equiv k_B T \Delta \mu$, which
is defined by the standard expression \cite{note_on_delta_mu}
\begin{equation}\label{def of delta mu}
\Delta \tilde{\mu} = k_B T  \ln \left( \frac{[ATP] \, [ADP]_{eq}
\, [P]_{eq}} {[ATP]_{eq}\, [ADP] \, [P]} \right),
\end{equation}
where $[..]$ denotes concentration under experimental conditions
and $[..]_{eq}$ denotes equilibrium concentrations. The chemical
potential of the hydrolysis reaction introduces a bias in the
dynamics of the motor, which is responsible for breaking the
time-reversal symmetry associated with the detailed balance
condition (which holds at equilibrium).

Following Ref.~\cite{kolomeisky}, the transition rates can be
generalized to include an external force $F_e$ according to
$\overleftarrow{\omega}_{i}^{\,l}(F_e) =
\overleftarrow{\omega}_{i}^{\,l}(0)\,e^{-\theta^{-}_i f}$ and
$\overrightarrow{\omega}_{i}^{\,l}(F_e) =
\overrightarrow{\omega}_{i}^{\,l}(0)\,e^{+\theta^{+}_i f}$, where
$f \equiv F_{e} d /( k_B T)$ and $\theta^{\pm}_i$ are the load
distribution factors. These load distribution factors take into
account the fact that the external force may not distribute
uniformly among different transitions \cite{hill}. Thus, we may
write the non-zero rates in the presence of force as:\bb
\begin{array}{ll}
\overleftarrow{\omega_b}^{-1} = \alpha\, e^{-\theta^{-}_b f}, & \overleftarrow{\omega_b}^0
 =  \omega\,e^{-\theta^{-}_b f},  \\
\overrightarrow{\omega_a}^1   =  \alpha\, e^{ -\epsilon + \Delta \mu + \theta^{+}_a f },
& \overrightarrow{\omega_a}^0 = \omega\,e^{-\epsilon + \theta^{+}_a f},    \\
\overleftarrow{\omega_a}^1  =  \alpha' e^{ -\epsilon + \Delta \mu - \theta^{-}_a f},
 & \overleftarrow{\omega_a}^0  = \omega'\,e^{-\epsilon- \theta^{-}_a f},\\
\overrightarrow{\omega_b}^{-1} = \alpha' e^{\theta^{+}_b f}, & \overrightarrow{\omega_b}^0
  =  \omega'\,e^{\theta^{+}_b f} \, .
\end{array}
\label{rates} \en  In the above expressions, the values of
 the parameters   $\theta_i^\pm$ are arbitrary
 except for the following constraint:
 After one period, the work done
by $F_e$ on the motor is $-F_e 2d$, implying that
$\theta^{+}_a+\theta^{-}_b + \theta^{-}_a+\theta^{+}_b\ =2$.
 Indeed, as shown in fig~\ref{fig:fits},  the simplest
 model with all the  $\theta_i^\pm$'s equal to
$1/2$, which was studied in  Ref.~\cite{nelson}, does  not
  reproduce the experimental curves of velocity versus force for
kinesin. The fact that the   $\theta_i^\pm$'s are different
  from 1/2   agrees with standard models of
kinesin, in which several chemical transitions are involved, and
the force must be split unequally among the different transition
rates \cite{howard}. We note  that this splitting of the force (which involves
the  actual value of the $\theta_i^\pm$'s) is a matter of kinetics, whereas  thermodynamics
enforces  only Eqs.~\ref{thermo_rates1}-\ref{thermo_rates2}. The expression  of
the rates given in  Eqs.~\ref{rates f=0}-\ref{rates} is essential for the
analysis which we develop below:  we  emphasize that these expressions
 are based on \emph{first principles}. Once these rates
are decomposed into an active and a passive part, the ratio of the
passive transition rates in Eq.~\ref{thermo_rates2} follows from
the condition of micro-reversibility (detailed balance), while the
ratio of the active transition rates in Eq.~\ref{thermo_rates1}
requires a more general principle for non-equilibrium chemical
reactions. Such a principle is based on the notion of affinity,
introduced by de Donder in Ref.~\cite{deDonder} to characterize
non-equilibrium chemical reactions. The de Donder equation relates
the forward and the backward reactions rates
$\overrightarrow{\alpha}$ and $\overleftarrow{\alpha}$ of an
elementary step, as a consequence of transition state theory:
\begin{equation}\label{De Donder}
\frac{\overrightarrow{\alpha}}{\overleftarrow{\alpha}}=e^{A/k_B
T},
\end{equation}
where $A$ is the affinity, defined as $-(\partial G/\partial
\zeta)_{T,P}$ in terms of the Gibbs free energy $G$ and $\zeta$
the extend of reaction. At equilibrium, $A=0$ and Eq.~\ref{De
Donder} leads to $\overrightarrow{\alpha}=\overleftarrow{\alpha}$,
which is the principle of micro-reversibility. Note that
Eq.~\ref{thermo_rates1} is indeed of the form of Eq.~\ref{De
Donder}, with the choice $\overrightarrow{\alpha}=\alpha_a$,
$\overleftarrow{\alpha}=\alpha_b$ and $\Delta \mu=A/ k_B T$ when
$\epsilon=0$. Thus we can consider Eq.~\ref{thermo_rates1} as a De
Donder relation, which generalizes the condition of
micro-reversibility far from equilibrium \cite{deDonder}.
Equivalently one can also interpret this equation as a particular
case of generalized steady state balance conditions, we shall come
back to this point in section IV.

\subsection{Effective description of the dynamics}
Let us now analyze further the conditions for directed motion for
this model, which as we mentioned earlier are required  to break
both the spatial symmetry and the symmetry associated with
detailed balance. These conditions for directed motion can be
derived by constructing an effective dynamics, which holds at long
times and large length scales \cite{nelson}. Let us first consider
a coarse-grained description in which the position variable $n$ is
the only state variable. The chemical variable $y$ may not be
accessible or we simply do not wish to include it in this
description. The dynamics of the motor is then described formally
by a master equation, which can be obtained from Eq.~\ref{Master
eq} by integrating out over all possible values of $y$. We are
then left with a coarse-grained master equation for $P_i(n,t)=\int
dy P_i(n,y,t)$, which is
\begin{eqnarray}
\partial_t P_i(n,t)  = -  \left( \overleftarrow{\omega}_i + \overrightarrow{\omega}_i
\right) P_{i}(n,t)+
\phantom{\overleftarrow{\omega}_{n+1}^{\,l}ppppppp}\nonumber \\
 \left [\,
\overleftarrow{\omega}_{j} \,P_{j}(n+1,t)+
\overrightarrow{\omega}_{j}\,P_{j}(n-1,t) \,\right ], \label{1 var
Master eq}
\end{eqnarray}
with the same rates as before. Note that these rates may still
depend  on the ATP concentration. As shown in Ref.~\cite{nelson},
an effective potential can be constructed for this problem by
eliminating one of the sites (a or b) from the master equation,
Eq.~\ref{1 var Master eq},  and describing the remaining dynamics
in terms of an effective potential. This is the effective
potential under which a random walker satisfying detailed balance
would exhibit the same dynamics. Of course, the same effective
evolution equation applies to occupation probabilities of site a
or b. This reasoning \cite{nelson} gives the effective energy
difference $\Delta E= E(n+2)-E(n)$ between site $n$ and site
$n+2$, which we write as $\Delta E= 2 k_B T \Psi$ with
\begin{equation}\label{Effective potential}
\Psi= \frac{1}{2} \ln \left( \frac{\Wl_a \Wl_b}{\Wr_a \Wr_b}
\right).
\end{equation}
When the rates of Eq.~\ref{rates} are used, we find that
\begin{equation}\label{Effective potential2}
\Psi= \frac{1}{2} \ln \left( \frac{ \left( \alpha+\omega \right)
\left( \alpha' e^{\Delta \mu} + \omega' \right)}{\left( \alpha
e^{\Delta \mu} +\omega \right) \left( \alpha'  + \omega' \right)}
\right) - f.
\end{equation}
Note that the effective potential is independent of the load
distribution factors $\theta_i^\pm$, and is identical to the
expression obtained in Ref.~\cite{nelson} except for the change in
the sign of the force \cite{notation}. A nice feature of
Eq.~\ref{Effective potential2} is that the conditions for
directional motion can now be immediately obtained from it, in a
way that is completely analogous to what is done for the ratchets
models in Ref.~\cite{armand1}. Directed motion is only possible if
the effective potential is tilted {\it i.e.},  $\Delta E \neq 0$.
Thus directed motion requires: (i) an asymmetric substrate which
means
 either $\alpha \neq \alpha'$ or $\omega \neq \omega'$, and (ii) breaking
of the detailed balance condition, so that  either  $\Delta \mu \neq 0$ or
$f \neq 0$. When $\Delta \mu=f=0$ the system is in
equilibrium, the effective potential is flat ($\Delta E=0$)  and
 no directional motion is possible. A difference between
this model and with the various ratchet models of Ref.~\cite{armand1}, is that
in ratchets  the position of the motor is  a
continuous variable. In the classification  of ratchet models
 given in~\cite{parmeggiani}, our model corresponds
to a system of class A  for which diffusion is not necessary for
motion generation. In this class of models, the two ratchet
potentials are identical and shifted with respect to each other in
such a way that each chemical cycle generates with a high
probability a step in the forward direction. As a consequence of
this construction, one should expect (and indeed we will find)
that in this model there is a strong coupling between the chemical
and mechanical coordinates, and the motor has a strong
directionality and a large thermodynamic efficiency.

\section{Modes of operation of the molecular motor, fit of experimental
curves and thermodynamic efficiency}

\subsection{Description of the dynamics using generating functions}

In this section, we analyze the long time behavior of our model
using generating functions, which has the additional advantage of
making the symmetry of the Fluctuation Theorem apparent as shown
in the next section. Let us introduce the generating functions:
$F_i(\lambda,\gamma,t) \equiv \sum_{y} \sum_{n} e^{-\lambda n -
\gamma y} P_{i}(n,y,t),$ whose time evolution is governed by:
$\partial_t F_i= {\cal M}_{ij}\,F_j$, where ${\cal M}[\lambda,
\gamma]$ is the following $2\times 2$ matrix which can be obtained
from the master equation of Eq.~\ref{Master eq}:
\begin{eqnarray} {\cal M}[\lambda, \gamma] =\left [
 \begin{array}{cc}
  -\overrightarrow{\omega_a}-\overleftarrow{\omega_a} &
  e^{\lambda}\, \Ul_b + e^{-\lambda}\,
  \Ur_b \\ e^{\lambda}\,
  \Ul_a
  +  e^{-\lambda}\,\Ur_a
   & -\overleftarrow{\omega_b}-\overrightarrow{\omega_b} \\
\end{array} \right ], \nonumber \\
\end{eqnarray}
with $\Ur_n(\gamma) \equiv \sum_{l} \overrightarrow{\omega_n}^l
e^{- l \gamma}$, and $\Ul_n(\gamma) \equiv \sum_{l}
\overleftarrow{\omega_n}^l e^{- l \gamma}$.

For $t \rightarrow \infty$,  we find
  \bb \left \langle\,e^{-\lambda n - \gamma y}\,\right
\rangle = \sum_{i} F_i(\lambda,\gamma,t) \sim \exp \left(
\vartheta\,t \right), \label{Evolution for F} \en  where $
\vartheta \equiv \vartheta[\lambda,\gamma]$ is the largest
eigenvalue of ${\cal M}$. This eigenvalue, $\vartheta$, contains
all the steady-state properties of the motor and its exact
expression is given by:
\begin{eqnarray} \vartheta(\lambda)= & \frac{1}{2} & \{ -\omega_a-\omega_b
+ [ (\omega_a - \omega_b )^2   \label{explicittheta_1}  \\
 & + & 4 ( \Ul_b e^\lambda + \Ur_b
e^{-\lambda} )  (  \Ul_a e^\lambda + \Ur_a e^{-\lambda} )
 ]^{1/2} \}, \nonumber
\end{eqnarray}
with the notations $\omega_a=\Wr_a+\Wl_a$ and
$\omega_b=\Wr_b+\Wl_b$.

The average (normalized) velocity  $\bar{v}$ is the current of the
mechanical variable, which is given by
\begin{equation}\label{def of v}
\bar{v} = \lim_{t \rightarrow \infty} \frac{<n(t)>}{t},
\end{equation}
and similarly the average ATP consumption rate $r$ is the current
of the chemical variable, which is given by
\begin{equation}\label{def of r}
r = \lim_{t \rightarrow \infty} \frac{<y(t)>}{t}.
\end{equation}
From Eq.~\ref{Evolution for F}, we see that $\bar{v} = -
{\partial_{\lambda}\vartheta }[\,0,0]$ and $r= -{
\partial_{\gamma} \vartheta }[\,0,0]$, and from Eq.~\ref{explicittheta_1}
we find explicitly that \bb \bar{v}=2 \frac{\Wr_a \Wr_b - \Wl_a
\Wl_b}{\Wr_a+\Wr_b+\Wl_a+\Wl_b}, \label{explicit v} \en \bb r=
\frac{\left( \Wl_a^1+\Wr_a^1 \right) \left( \Wr_b+\Wl_b \right) -
\left( \Wl_b^{-1}+\Wr_b^{-1} \right) \left( \Wr_a+\Wl_a \right)}
{\Wr_a+\Wr_b+\Wl_a+\Wl_b} \, . \label{explicit r} \en The method
gives also access to higher moments of $n(t)$ and $y(t)$. The
second moments for instance can be expressed in terms of the
diffusion matrix
\begin{equation}\label{Diffusion matrix}
2 D_{ij}=  \frac{\partial^2 \vartheta}{\partial z_i \partial
z_j}\, [\,0,0 ],
\end{equation}
with the understanding that $z_1=\lambda$ and $z_2=\gamma$. These
first and second moments can also be obtained by calculating the
average of $n(t)$ and $y(t)$ directly from the master equation
Eq.~\ref{Master eq} \cite{nelson,kolomeisky,derrida}.

\subsection{Modes of operation of the motor}
\begin{figure}
\includegraphics[height=1.9in]{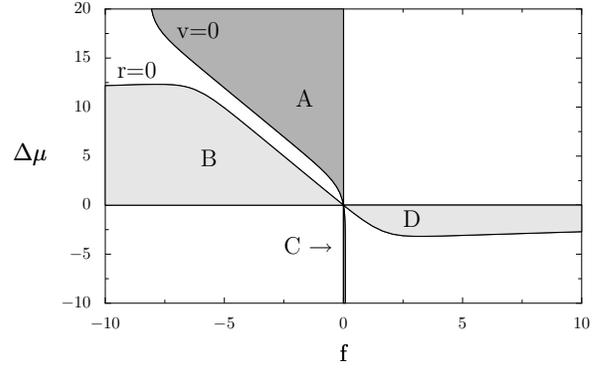}
\caption{Four modes of operation of a molecular motor, as
delimited by $\bar{v}=0$ and $r=0$ \cite{armand1}.  The lines are
generated with parameters that we have extracted by  fitting the
data for kinesin in Ref.~\cite{block} to our model, and this fit
is shown in fig.~\ref{fig:fits}. } \label{fig:modes_operation}
\end{figure}
From the conditions of vanishing of the currents: $\bar{v}=0$ and
$r=0$, we can construct a full operation diagram of a motor, as
shown in Fig.\ \ref{fig:modes_operation} for the case of kinesin.
The curves $\bar{v}=0$ and $r=0$ define implicitly $f =
f_{\mbox{\scriptsize st}}(\Delta \mu)$ (the stalling force) and
$\Delta \mu = \Delta \mu_{\mbox{\scriptsize st}}(f)$,
respectively. The stalling force is
\begin{equation}\label{Stalling force}
f_{\mbox{\scriptsize st}}(\Delta \mu)=\frac{1}{2} \ln \left(
\frac{ \left( \alpha+\omega \right) \left( \alpha' e^{\Delta \mu}
+ \omega' \right)}{\left( \alpha e^{\Delta \mu} +\omega \right)
\left( \alpha'  + \omega' \right)} \right),
\end{equation}
which means that for $f=f_{\mbox{\scriptsize st}}(\Delta \mu)$,
$v=0$ and $\Psi=0$, where  $\Psi$ was defined in
Eq.~\ref{Effective potential2}. At the stalling  point,  the
mechanical variable is equilibrated but not the chemical variable.
Therefore, in general, the motor consumes  ATP {\it i.e.},  $r
\neq 0$,  even if it
  is stalled (in fact, it is only at equilibrium,
 $f=\Delta \mu=0$,  that both  $v$ and  $r$   vanish).
Near stalling for $\Delta \mu \neq 0$, the motor evolves in a
quasi-static manner but irreversibly. A similar phenomenon occurs
in thermal ratchets \cite{ken,parrondo}.

  Likewise,  the condition $\Delta
\mu=\Delta \mu_{\mbox{\scriptsize st}}(f)$ means that $r=0$.
 The explicit form of $\Delta
\mu_{\mbox{\scriptsize st}}(f)$ is \bb \Delta
\mu_{\mbox{\scriptsize st}}(f)= \ln \frac{ \left(\alpha
e^{-\theta_b^- f}+\alpha' e^{\theta_b^+ f} \right) \left(\omega
e^{\theta_a^+ f} + \omega' e^{-\theta_a^- f} \right)}{\left(\alpha
e^{\theta_a^+ f}+\alpha' e^{-\theta_a^- f} \right) \left(\omega
e^{-\theta_b^- f} + \omega' e^{-\theta_b^+ f} \right)}.  \en  The
four different regimes of operation of the motor, discussed in
Refs.~\cite{armand1,prl} for ratchet models can be recovered here.
In Region A, where $r \Delta \mu
>0$ and $f \bar{v}<0$, the motor uses chemical energy of ATP to
perform mechanical work. This can be understood by considering a
point on the y-axis of Fig.\ \ref{fig:modes_operation} with
$\Delta \mu>0$. There we expect that the motor drifts to the right
with $\bar{v}>0$. Now in the presence of a small load $f<0$ on the
motor, we expect that the motor is still going in the same
direction although the drift is uphill and thus work is performed
by the motor at a rate $\dot{W}=-f\bar{v} >0$. This holds as long
as $f$ is smaller than the stalling force, which defines the other
boundary of region A. Similarly, in Region B, where $r \Delta \mu
<0$ and $f \bar{v}>0$, the motor produces ATP from mechanical
work. In Region C, where $r \Delta \mu >0$ and $f \bar{v} <0$, the
motor uses ADP to perform mechanical work. In Region D, where $r
\Delta \mu <0$ and $f \bar{v}>0$, the motor produces ADP from
mechanical work. It is interesting to note that the large
asymmetry between regions A and C in Fig.\
\ref{fig:modes_operation} reflects the fact that kinesin is a
unidirectional motor. Furthermore the regions A and B do not touch
except at the origin. With kinesin operating in normal conditions
in region A with $\Delta \mu \simeq 15$, the presence of a gap
between regions A and B means that kinesin should not be able to
switch into an ATP producing unit (region B), and indeed this has
never been observed experimentally. Note that the explicit
expressions for $f_{\mbox{\scriptsize st}}$ and $\Delta
\mu_{\mbox{\scriptsize st}}$ obtained in this model do not depend
on the energy difference  $\epsilon$ between the two states,  due
to a cancellation of  the numerator and denominator in
Eqs.~\ref{explicit v}-\ref{explicit r}. Thus the diagram of
operation of the motor is valid for arbitrary value of $\epsilon$.

\subsection{Fit of experimental curves of velocity versus force for kinesin}
\begin{figure}
\includegraphics[height=1.9in]{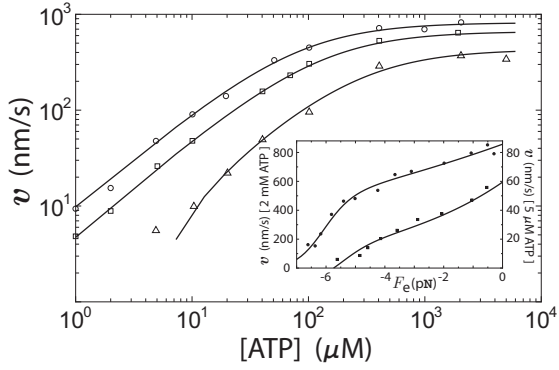}
\caption{Kinesin velocity vs.\ ATP concentration under an external
force \cite{prl}. The solid curves are the fits of our model to
data from Ref.\ \cite{block}. From the top down, the plots are for
$F_e = -1.05, -3.59$, and $-5.63\,\mbox{pN}$, respectively. Inset:
Kinesin velocity vs.\ force under a fixed ATP concentration.  The
solid curves are fits to the data of Ref.\ \cite{block}. From the
top down, the plots are for $[\mbox{ATP}] = 2\,\mbox{mM}$ and
$5\,\mbox{$\mu$M}$.  From this fit, we obtained the following
parameters for our model : $\epsilon = 10.81$, $k_0 = 1.4 \cdot
10^{5}\,\mbox{$\mu$M}^{-1}$, $\alpha = 0.57\,\mbox{s}^{-1}$,
$\alpha' = 1.3 \cdot 10^{-6}\,\mbox{s}^{-1}$, $\omega =
3.5\,\mbox{s}^{-1}$, $\omega'= 108.15\,\mbox{s}^{-1}$,
$\theta^{+}_a = 0.25$, $\theta^{-}_a = 1.83$, $\theta^{+}_b=0.08$,
and $\theta^{-}_b = -0.16$.  } \label{fig:fits}
\end{figure}

  We  now discuss how the parameters of the model were determined
using experimental data obtained for kinesin. In
fig.~\ref{fig:fits} (which is also fig.~4 of Ref.~\cite{prl}), we
have fitted velocity vs.\ force curves for two values of ATP
concentrations, and also several curves of velocity vs.\ ATP
concentration at different forces using the data of
Ref.~\cite{block}. We have assumed that $e^{\Delta \mu} =
k_0\,[\mbox{ATP}]$, which is well verified at moderate or high
levels of ATP. At low concentration of ATP, there is no such
simple correspondence because it is no more  legitimate  to treat
the ADP and P concentrations as constant. We think that this is
probably the reason why the fit is not as  good for the lowest ATP
values (this concerns the first point in the curve at $F_e=-5.63$
pN and low ATP value in fig.~\ref{fig:fits}). Nevertheless, we can
fit very well the majority of the experimental points with this
model and we obtained the following values for the parameters:
$\epsilon = 10.81$, $k_0 = 1.4 \cdot 10^{5}\,\mbox{$\mu$M}^{-1}$,
$\alpha = 0.57\,\mbox{s}^{-1}$, $\alpha' = 1.3 \cdot
10^{-6}\,\mbox{s}^{-1}$, $\omega = 3.5\,\mbox{s}^{-1}$, $\omega'=
108.15\,\mbox{s}^{-1}$, $\theta^{+}_a = 0.25$, $\theta^{-}_a =
1.83$, $\theta^{+}_b=0.08$, and $\theta^{-}_b = -0.16$.  These
values are reasonable within the present accepted picture of the
nano-operation of kinesin \cite{howard}. Indeed,   $\epsilon$ and
$k_0^{-1}$  represent, respectively,  the typical binding energy
($\sim 10\,k_B T$) of kinesin with microtubules and the ATP
concentration at equilibrium ($\sim 10^{-5}\,\mbox{$\mu$M}$).
 Moreover, $\theta^{-}_a = 1.83$ indicates that the back-steps
(transitions $a \rightarrow b$) of kinesin contain most of the
force sensitivity \cite{howard}. Furthermore, our framework allows
us to estimate a maximum stalling force of $-7\,\mbox{pN}$.

A useful quantity to consider is the distance $\ell$, which the
motor walk using the hydrolysis of one ATP molecule, which is
\begin{equation}\label{ell}
\ell = \frac{\bar{v}}{r}.
\end{equation} We find for this model in agreement with Ref.~\cite{block}
that $\ell \simeq 0.97 (2d)$ in the absence of load, which
corresponds to one step (8nm) per hydrolysis of one ATP molecule.
Thus the coupling ratio of kinesin is indeed independent of ATP
concentration and is 1:1 at negligible loads. We also find a
global ATP consumption rate of $r \simeq 111\,\mbox{s}^{-1}$, in
excellent agreement with known values \cite{howard}. It should be
remarked that the global ATP consumption rate measurements done in
ATPase assays in the bulk are in agreement with the single
molecule experiments, which are intrinsically very different
experiments. Kinesin is well described by tightly coupled models
which incorporate a single mechanically sensitive rate and this is
consistent with our findings that there is only one transition
(transitions $a \rightarrow b$) that  has all the force
sensitivity {\it i.e.}, the largest load distribution factor. In
principle,  by changing the parameters of the model, we could
characterize motors which are less tightly coupled, but there will
always be some coupling because, by construction,   the
 mechanical steps are intrinsically  linked with the chemical cycle.

We have compared our fit with that carried out by Fisher et al. in
Ref.~\cite{kolomeisky} where the same data was fitted, and we
observe  that the outcome of both fits is comparable. In this
comparison, there is an issue of complexity of the model under
consideration to be taken into account. This is especially
important in fitting experimental data of kinetics, which is
typically hard to fit because one has many parameters to fit in an
expression which is a sum of exponential functions. The model of
Ref.~\cite{kolomeisky} is of higher complexity because it involves
4 states instead of 2 states for our model, thus we might be
tempted to say that our model does better in fitting the same data
with less complexity. We believe that this is true when
considering the data for the velocity only, but if we were to
include also the data for the diffusion coefficient (which is
related to the randomness parameter defined in Ref.~\cite{block}),
we agree with Ref.~\cite{kolomeisky} that a model with 4 states
would then do better than a model with only 2 states.

\subsection{Thermodynamic Efficiency}
\begin{figure}
\includegraphics[height=1.9in]{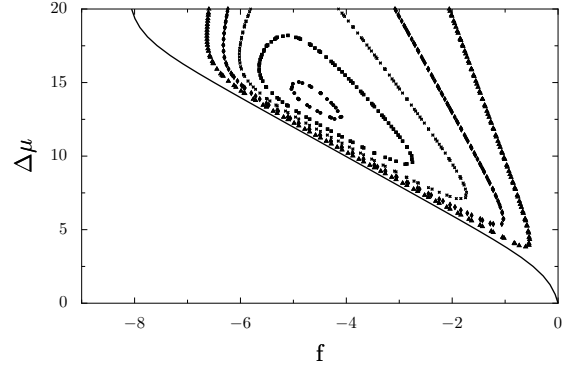}
\caption{Curves of equal efficiency $\eta$ within region A (which
is delimited by the solid line and by the $y$ axis). The
parameters are those used in fig.~\ref{fig:modes_operation} and
obtained from the fit of
 fig.~\ref{fig:fits}. From the outside to the inside the curves
correspond to   $\eta=0.2$,
  $\eta=0.3$,   $\eta=0.4$, $\eta=0.5$ and $\eta=0.58$.
 The absolute maximum efficiency for these
parameters is about 59\% and is located at $\Delta \mu \simeq 14$
and $f \simeq -4.9$. } \label{fig:efficiency}
\end{figure}

Another important quantity that characterizes the working of a
motor is its efficiency \cite{ken,armand1}.  In region A, it is
defined as the ratio of the work performed to the chemical energy:
\bb \eta = - \frac{f \bar{v}}{r \Delta \mu}= -\frac{f \ell}{\Delta
\mu}. \en By definition, $\eta$ vanishes at $f=0$ and at the
stalling force $f_{\mbox{\scriptsize st}}(\Delta \mu)$. Therefore,
it has a local maximum $\eta_{\mbox{\scriptsize m}}(\Delta \mu)$
for some $f_{\mbox{\scriptsize m}}(\Delta \mu)$ between
$f_{\mbox{\scriptsize st}} < f_{\mbox{\scriptsize m}} < 0$. Near
equilibrium, $\eta_{\mbox{\scriptsize m}}(\Delta \mu)$ has a
constant value, $\eta_{\mbox{\scriptsize m}}^{\mbox{\scriptsize
eq}}$, along a straight line $f_{\mbox{\scriptsize m}}(\Delta \mu)
\propto \Delta \mu$ inside region A \cite{armand1}. However, far
from equilibrium, the picture is drastically different. We find
that (i) $f_{\mbox{\scriptsize m}}(\Delta \mu)$ is no longer a
straight line, (ii) $\eta_{\mbox{\scriptsize m}} -
\eta_{\mbox{\scriptsize m}}^{\mbox{\scriptsize eq}} \propto \Delta
\mu$ for small $\Delta \mu$, and (iii) $\eta_{\mbox{\scriptsize
m}} \sim 1/\Delta \mu$ for large $\Delta \mu$. Therefore,
$\eta_{\mbox{\scriptsize m}}$ must have an absolute maximum at
some $\Delta \mu > 1$. One can also consider the curves of equal
value of the efficiency within region A. In the particular case of
the linear regime close to equilibrium, these curves are straight
lines going through the origin \cite{armand1}, but in general far
from equilibrium these curves are not straight lines as can be
seen in Fig.~\ref{fig:efficiency}, and the maximum efficiency is
reached at a point within region A.

Note that $\eta_{\mbox{\scriptsize m}}$ is substantially larger
than $\eta_{\mbox{\scriptsize m}}^{\mbox{\scriptsize eq}}$. For
instance with the parameters used in fig.~\ref{fig:efficiency},
the maximum efficiency is around 0.59 while
$\eta_{\mbox{\scriptsize m}}^{\mbox{\scriptsize eq}} \simeq 0.03$
(see also Fig. 3b of Ref.~\cite{prl} which contains a plot of
$\eta_{\mbox{\scriptsize m}}$ as a function of $\Delta \mu$ under
the same conditions). Hence, this motor achieves a higher
efficiency in the far-from-equilibrium regime as was also found in
other studies of molecular motors using continuous ratchet models
(see e.g., Ref.~\cite{parmeggiani}). Under typical physiological
conditions ($\Delta \widetilde{\mu} \sim 10 - 25\,\mbox{$k_B
T$}$), kinesin operates at an efficiency in the range of $40 -
60\%$,  in agreement with experiments \cite{howard}. It is
interesting to note that kinesins operate most efficiently in an
energy scale corresponding to the energy available from ATP
hydrolysis.

\section{Finite time and long time fluctuation theorem}
\subsection{Long time FT}

We note that the rates of Eq.~\ref{rates} satisfy the following
generalized detailed balance conditions:
\begin{eqnarray} \overrightarrow{\omega_b}^{-l}
P_b^{\mbox{\scriptsize eq}} & = & \overleftarrow{\omega_a}^{l}
P_a^{\mbox{\scriptsize eq}}\,e^{+(\,\theta^{-}_a+\theta^{+}_b\,)f
- \Delta \mu\,l},
\label{Generalized db1}\\
\overleftarrow{\omega_b}^{-l}\,P_b^{\mbox{\scriptsize eq}} & = &
\overrightarrow{\omega_a}^{l} P_a^{\mbox{\scriptsize eq}}\, e^{-
(\,\theta^{+}_a+\theta^{-}_b\,)f - \Delta \mu\,l},
\label{Generalized db2}
\end{eqnarray}
for $l=0,1$.  Here,  $P_a^{\mbox{\scriptsize eq}} =
1/(1+e^{-\epsilon})$ and $P_b^{\mbox{\scriptsize eq}} =
e^{-\epsilon}/(1+e^{-\epsilon})$ are the equilibrium probabilities
corresponding to $f=0$ and $\Delta \mu =0$. We note that these
relations, Eq.~\ref{Generalized db1}  and   Eq.~\ref{Generalized
db2}, while valid arbitrarily far from equilibrium, still refer to
the equilibrium state via the probabilities $P_i^{eq}$. Using the
definition of the equilibrium probabilities, one can in fact
rewrite Eqs.~\ref{Generalized db1}-\ref{Generalized db2} as
\begin{eqnarray}
\ln\frac{\overrightarrow{\omega_b}^{-l}}{\overleftarrow{\omega_a}^{l}}
 & = &
\epsilon + \left(\theta^{-}_a+\theta^{+}_b\, \right) f - \Delta
\mu\,l,
\label{Generalized db1b}\\
\ln\frac{\overleftarrow{\omega_b}^{-l'}}{\overrightarrow{\omega_a}^{l'}}
 & = &
\epsilon - \left(\theta^{+}_a+\theta^{-}_b\, \right) f - \Delta
\mu\,l', \label{Generalized db2b}
\end{eqnarray}
for $l,l'=0,1$. Note that these relations are analogous to the De
Donder relation of Eq.~\ref{De Donder} and to the transition state
theory equations of Eqs.~\ref{thermo_rates1}-\ref{thermo_rates2}.
Moreover, by combining these two equations, using the constraint
that the sum of the load distribution factors is two and then
multiplying the result by $k_B T$, one obtains
\begin{equation}\label{Steady state balance condition}
k_B T \ln\frac{\overrightarrow{\omega_b}^{-l}
\overrightarrow{\omega_a}^{l'}}{\overleftarrow{\omega_a}^{l}
\overleftarrow{\omega_b}^{-l'}}= F_e (2d) - \Delta \tilde{\mu}
\left( l-l' \right),
\end{equation}
which has the form of the steady state balance condition discussed
in Refs~\cite{liepelt,seifert_entropy}. As pointed out in these
references, by identifying the left hand side of Eq.~\ref{Steady
state balance condition} with the heat delivered to the medium,
i.e. with the change of entropy of the medium, the right hand side
of Eq.~\ref{Steady state balance condition} can be interpreted as
the sum of the mechanical work $F_e (2d)$ and the chemical work $-
\Delta \tilde{\mu} \left( l-l' \right)$ on that particular set of
cyclic transitions $(l,l')$. In that sense, Eqs.~\ref{Generalized
db1b}-\ref{Steady state balance condition} can be understood as
formulations of the first law at the level of elementary
transitions. It is interesting to see that these steady state
balance relations also lead to a FT as we now show below.

\begin{figure}
\includegraphics[height=1.9in]{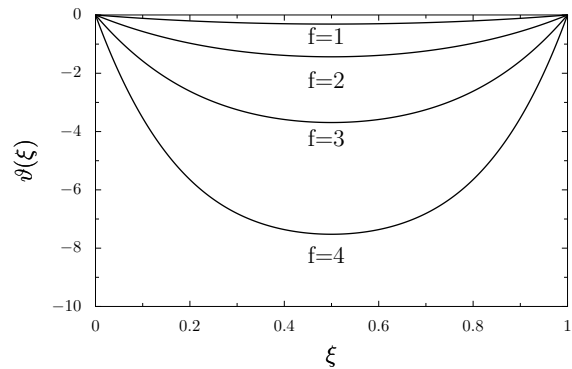}
\caption{Graphical illustration of the symmetry of the long-time
Fluctuation Theorem (for simplicity a model without chemical
variable has been used). The largest eigenvalue $\vartheta$ is
shown as function of $\xi=\lambda/f$ for different values of the
normalized force $f$. The symmetry of the long time Fluctuation
Theorem corresponds to the symmetry of this curve with respect to
$\xi=1/2$.} \label{fig:long time FT}
\end{figure}

 Using Eqs.\ (\ref{Generalized db1})
and (\ref{Generalized db2}), it can be shown that ${\cal M}$
  and ${\cal M}^{\dag}$, the adjoint of ${\cal M}$,  are
related by a similarity transformation:
\begin{equation}
{\cal M}^{\dag}[\,f  - \lambda , \Delta \mu - \gamma ]=  {\cal
Q}\,{\cal M}[\lambda , \gamma]\,{\cal Q}^{-1}, \label{similarity
tf} \end{equation} where ${\cal Q}$ is the following diagonal matrix:
\begin{eqnarray} {\cal Q} =\left [
 \begin{array}{cc}
P_b^{eq} e^{(\theta_a^+ + \theta_b^-) f/2} &  0 \\
0  & P_a^{eq} e^{(\theta_a^- + \theta_b^+) f/2}\\
\end{array} \right ]. \label{matrice Q}
\end{eqnarray}
This similarity relation implies that ${\cal M}[\lambda , \gamma]$
and ${\cal M}^{\dag}[\,f  - \lambda , \Delta \mu - \gamma ]$ have
the same spectra of eigenvalues and therefore \bb
\vartheta[\lambda,\gamma]= \vartheta[\,f -\lambda,\Delta \mu -
\gamma], \label{FT} \en which is one form of FT. Since this
relation holds at long times irrespective of the initial state, it
is a Gallavotti-Cohen relation \cite{FT}. Such a symmetry is
illustrated graphically on Fig.~\ref{fig:long time FT}  for  a
simplified case where the chemical variable is absent.

\subsection{Implications of FT in the linear regime}
Here, we discuss the implications of FT in the linear regime,
which leads to the Einstein and Onsager relations near
equilibrium. Differentiating Eq.~\ref{FT} with respect to
$\lambda$ and $\gamma$, we obtain:
\begin{eqnarray} \bar{v} = -  \frac{\partial \vartheta}{\partial \lambda}
[\,0,0] & = & \frac{\partial \vartheta}{\partial \lambda}[\,f,
\Delta \mu], \label{v and
r1} \\
r = - \frac{\partial \vartheta}{\partial \gamma} [\,0,0] & = &
\frac{\partial \vartheta}{\partial \gamma} [\,f,\Delta \mu].
\label{v and r2}
\end{eqnarray}
The response and fluctuations of a motor are quantified,
respectively, by a response matrix $\lambda_{ij}$ and by the
diffusion matrix $D_{ij}$ defined in Eq.~\ref{Diffusion matrix}.
The physical meanings of $\lambda_{ij}$ are: $\lambda_{11} \equiv
\partial \bar{v} /\partial f$  is the mobility, $\lambda_{22}
\equiv
\partial r /\partial \Delta \mu$ is the chemical admittance, and
 $\lambda_{12} \equiv \partial \bar{v} /\partial
\Delta \mu$ and $\lambda_{21} \equiv \partial r/\partial f$ are
the Onsager coefficients that quantify the mechanochemical
couplings of the motor. Near equilibrium, where $f$ and $\Delta
\mu$ are small, a Taylor expansion of the r.h.s. of Eq.~\ref{v and
r1} and Eq.~\ref{v and r2} with respect to $f$ and $\Delta \mu$
leads to
\begin{eqnarray} \left. \frac{\partial \vartheta}{\partial \lambda} \right|_{\,f,
\Delta \mu} & \simeq & \left. \frac{\partial \vartheta}{\partial
\lambda}\right|_{0,0} + f \left. \frac{\partial^2
\vartheta}{\partial^2 \lambda}\right|_{0,0} + \Delta \mu \left.
\frac{\partial^2
\vartheta}{\partial \lambda \partial \gamma}\right|_{0,0}, \label{Taylor1}  \\
\left. \frac{\partial \vartheta}{\partial \gamma} \right|_{\,f,
\Delta \mu} & \simeq & \left. \frac{\partial \vartheta}{\partial
\gamma}\right|_{0,0} + f \left. \frac{\partial^2
\vartheta}{\partial \gamma \partial \lambda}\right|_{0,0} + \Delta
\mu \left. \frac{\partial^2 \vartheta}{\partial^2
\gamma}\right|_{0,0}. \label{Taylor2}
\end{eqnarray}
Using the definitions of $\bar{v}$ and $r$ from Eqs.~\ref{v and
r1}-\ref{v and r2}, one obtains directly
\begin{eqnarray} \bar{v}  & = & \lambda_{11}^0\,f +
\lambda_{12}^0\,\Delta \mu, \\  r & = & \lambda_{21}^0\,f +
\lambda_{22}^0\,\Delta \mu, \nonumber
\end{eqnarray} with $\lambda_{ij}^0 =
\partial_{z_i}\partial_{z_j} \vartheta [\,0,0]/2 \equiv D_{ij}$,
which are the Einstein relations, and $\lambda_{12}^0 \equiv
\partial_{\gamma}\partial_{\lambda} \vartheta [\,0,0]/2 =  \partial_{\lambda}
\partial_{\gamma}\vartheta [\,0,0]/2 \equiv \lambda_{21}^0$, which is
the Onsager relation.  Thus, FT describes the response and
fluctuations near equilibrium \cite{gallavotti,gaspard}.

It is interesting to investigate how Einstein or Onsager relations
are broken in non-equilibrium situations. The "violations" of
Einstein and Onsager relations when linear response theory is used
in the vicinity of a non-equilibrium state rather than near an
equilibrium state were studied in Ref.~\cite{prl}. There, we
quantified the violations of Einstein and Onsager relations,
respectively, by four temperature-like parameters, $T_{ij}$, and
by the difference of the mechanochemical coupling coefficients,
$\Delta \lambda$. Of course, these effective temperatures are not
thermodynamic temperatures: they are merely one of the ways to
quantify deviations from Einstein relations; similarly our
definition of $\Delta \lambda$ is just one of the possible ways to
study the "violations" of Onsager relations: strictly speaking
there are no real violations since Einstein and Onsager relations
apply only to systems at equilibrium. We have shown in~\cite{prl}
some of the possible behaviors of $T_{ij}$ and $\Delta \lambda$
for a kinesin motor using the parameters of the fit discussed
above:  in particular, we found  that for kinesin the maximum
value of $\Delta \lambda$ is   $\Delta \lambda \sim
45\,\mbox{pN}^{-1}\mbox{s}^{-1}$, and that
 at large $\Delta \mu$,  $\Delta \lambda \sim -10\,\mbox{pN}^{-1}\mbox{s}^{-1}$.
  We  also found that (i) one of the Einstein relations
holds near stalling (a point which we justify more precisely in
the next section in Eq.~\ref{Einstein rel stalling}), (ii) the
degree by which the Onsager symmetry is broken ($\Delta \lambda
\neq 0$) is largely determined by the underlying asymmetry of the
substrate, (iii) only two ``effective" temperatures characterize
the fluctuations of tightly coupled motors, (iv) kinesin's maximum
efficiency and the maximum violation of Onsager symmetry occur
roughly at the same energy scale, corresponding to that of ATP
hydrolysis ($\sim 20\,k_B T$) \cite{prl}. Experimental and
theoretical violations of the Fluctuation-Dissipation relation
have been observed and studied in many active biological systems
\cite{martin,jb_PRE,joanny,fred,andy}, but to our knowledge no
experiments testing the Fluctuation-Dissipation or the Onsager
relations have been carried out at the single motor level.

\subsection{Finite time FT}
\begin{figure}
\includegraphics[height=1.9in]{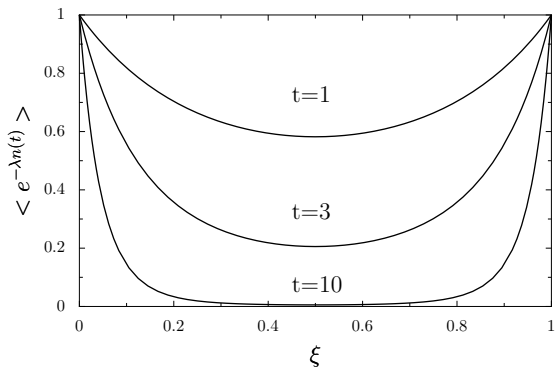}
\caption{Graphical illustration of the symmetry of the finite-time
Fluctuation Theorem for  a simplified  model without chemical
variable.  The left hand side of Eq.~\ref{FT finite time} is shown
as function of $\xi=\lambda/f$ for different times (arbitrary
units). The symmetry of the finite time Fluctuation Theorem
amounts to the symmetry of this curve with respect to $\xi=1/2$.}
\label{fig:finite time FT}
\end{figure}
The similarity transformation~(\ref{similarity tf})   implies that
{\it all} the eigenvalues of ${\cal M}[\lambda , \gamma]$ and
${\cal M}^{\dag}[\,f - \lambda , \Delta \mu - \gamma ]$ are
identical, not just the largest one. This more general property
allows us to prove a transient FT, because the dynamics of the
model at finite time involves all the eigenvalues of ${\cal M}$
and not just the largest one. The price to pay to have a FT
relation valid at finite time is that the initial state can no
more be arbitrary.
 We show here that the relation still holds, in the
particular case when the initial state is prepared to be in an
\emph{equilibrium state} (which corresponds to the condition
$f=\Delta \mu=0$) and when, in addition,  a specific condition on the
load distribution factors is obeyed. To see how this comes about
in this model, we  assume that the motor  at time $t=0$ is  at
the origin $n(0)=y(0)=0$ in an equilibrium state, and we calculate the
 values of the  position $n(t)$ and of the chemical variable
$y(t)$ at  time $t$. We  denote the initial state by the
vector
\begin{eqnarray} |F_0 \rangle = | F( t=0) \rangle=\left (
 \begin{array}{c}
  P_a^{eq}   \\
    P_b^{eq}  \\
\end{array} \right ). \nonumber
\end{eqnarray}
With $\langle 0 |=(1,1)$, the initial state vector is normalized
since $\langle 0 | F_0 \rangle=1$. Let us introduce ${\cal
U}(\lambda,\gamma,t)=e^{{\cal M}[\lambda, \gamma]t}$, the
evolution operator for the generating functions $F_i$. By taking the
exponential of  Eq.~\ref{similarity tf}, one finds that this
operator also obeys an FT relation
\begin{equation}
{\cal U}^{\dag}[\,f  - \lambda , \Delta \mu - \gamma,t ]=  {\cal
Q}\,{\cal U}[\lambda , \gamma,t]\,{\cal Q}^{-1}. \label{similarity
tf for U} \end{equation} We calculate  the following average,  similar  to
Eq.~\ref{Evolution for F},
\begin{eqnarray} \langle e^{-(f-\lambda) n(t)-(\Delta \mu -\gamma)
y(t)} \rangle
  & = & \langle 0 | {\cal U}(f-\lambda,\Delta \mu-\gamma,t) | F_0 \rangle,  \nonumber \\
  & = & \langle F_0 | {\cal U}^\dag(f-\lambda,\Delta \mu-\gamma,t) | 0 \rangle, \nonumber \\
  & = & \langle F_0 | {\cal Q} {\cal U}(\lambda,\gamma,t) {\cal Q}^{-1} | 0 \rangle, \nonumber \\
  & = & \langle 0 | {\cal U}(f,\gamma,t) | F_0 \rangle, \label{FT finite time} \\
  & = & \langle e^{-\lambda n(t)-\gamma y(t)} \rangle.
\nonumber
\end{eqnarray}
  We have used Eq.~\ref{similarity tf for U} to derive the
third equality, and  the final equation requires the condition:
$\langle F_0 |{\cal Q}= \langle 0 |$, which is equivalent to
${\cal Q}^{-1} | 0 \rangle = | F_0 \rangle$ since ${\cal Q}$ is
diagonal. Using Eq.~\ref{matrice Q}, we find that this relation
holds if  the initial state is in equilibrium and if  the
following condition holds \bb \theta_a^- + \theta_b^+=\theta_a^+ +
\theta_b^-. \en

 Equation~\ref{FT finite time} is analogous to the Evans
transient time Fluctuation Theorem \cite{evans} and to the Crooks
relation \cite{crooks}. An important point here is that the
initial state must be an equilibrium state while the final state
at time $t$ does not have to be  (and in general is not) an
equilibrium state. Crooks relation can be derived  using a path
representation of the ratio of forward to backward probabilities
according to a specific protocol, assuming a Markov process and
using a generalized detailed balance relation between successive
states. In our case, the equivalent of the generalized "local"
detailed balance condition needed for the proof is
Eq.~\ref{similarity tf for U}. The symmetry of the transient
Fluctuation Theorem is illustrated graphically in
Fig.~\ref{fig:finite time FT} for a simplified case where the
chemical variable is absent (or integrated out).

\section{Fluctuation Theorem for the large deviation function}

\subsection{Explicit calculation of the large deviation function of the current}
Here, we   again take  advantage of our knowledge of the function
$\vartheta$, which contains all the information about the long
time dynamical properties of the model, to obtain an explicit
expression  for the large deviation function of the current. To
simplify the presentation, we  consider the simplified
description, given in  Eq.~\ref{1 var Master eq}, in which the
chemical variable $y$ is not taken into account. In this case, the
generating function is defined by $F_i(\lambda,t) \equiv \sum_{y}
\sum_{n} e^{-\lambda n} P_{i}(n,y,t),$ and the matrix ${\cal
M}[\lambda]$ becomes
\begin{eqnarray} \label{Matrice M(lambda)} {\cal M}[\lambda] =\left [
 \begin{array}{cc}
  -\overrightarrow{\omega_a}-\overleftarrow{\omega_a} &
  e^{\lambda}\, \overleftarrow{\omega_b} + e^{-\lambda}\,
  \overrightarrow{\omega_b}\\ e^{\lambda}\,
  \overleftarrow{\omega_a}
  +  e^{-\lambda}\,\overrightarrow{\omega_a}
   & -\overleftarrow{\omega_b}-\overrightarrow{\omega_b} \\
\end{array} \right ]. \nonumber \\
\end{eqnarray}
By definition, $\vartheta$ is the largest eigenvalue of ${\cal
M}(\lambda)$, so similarly to Eq.~\ref{explicittheta_1} we have
\begin{eqnarray} \vartheta(\lambda)= & \frac{1}{2} & \{ -\omega_a-\omega_b
+ [ (\omega_a - \omega_b )^2   \label{explicittheta}  \\
 & + & 4 ( \Wl_b e^\lambda + \Wr_b
e^{-\lambda} )  (  \Wl_a e^\lambda + \Wr_a e^{-\lambda} )
 ]^{1/2} \}, \nonumber
\end{eqnarray}
with the notations $\omega_a=\Wr_a+\Wl_a$ and
$\omega_b=\Wr_b+\Wl_b$.

We have already seen that $\vartheta(\lambda)$ has the property
that $<e^{-\lambda n}> \sim e^{\vartheta(\lambda)t}$ for large
$t$. On the other hand, the large deviation function $G(v)$ is
defined for large time $t$ by
\begin{equation}\label{LD_def}
\mathcal{P}(\frac{n}{t}=v) \sim e^{-G(v)t},
\end{equation}
in terms of $\mathcal{P}(n/t=v)$ the probability to observe a
current $v$ after the motor has gone a distance $n$ from the
origin in a time $t$. The relation between $\vartheta(\lambda)$
and $G(v)$ is
\begin{eqnarray}
<e^{-\lambda n}> & = & \int e^{-\lambda n} dn \mathcal{P}(n), \\
       & = & \int t dv \mathcal{P}(\frac{n}{t}=v) e^{-\lambda v t}, \\
       & \sim & \int dv e^{\left(-G(v)-\lambda v \right)t}.
\end{eqnarray}
Using the saddle point method, we find that
$\vartheta(\lambda)=\max_v(-G(v)-\lambda v)$ and thus
$\vartheta(\lambda)$ and $G(v)$ are Legendre transform of each
other. We have also $-G(v)=\max(\vartheta(\lambda)+ \lambda v)$,
which can be written in parametric form
\begin{eqnarray}
 \frac{\partial \vartheta}{\partial \lambda}[\lambda=\lambda^*]
+v & = &
 0,  . \label{formeparamLD} \\
\vartheta(\lambda^*)+\lambda^* v  & = & -G(v) . \label{explicit
LD}
\end{eqnarray}
Using Eqs.~\ref{explicittheta}-\ref{explicit LD}, we find the
following expressions  for $G(v)$ (see Appendix A for details of
the derivation): for $v>0$
\begin{equation}\label{G(v)+}
G(v)  =  \frac{\omega_a+\omega_b}{2}+ \frac{\sqrt{\Omega}}{2v}
\left(Y^-(v) - \frac{1}{Y^-(v)} \right) - v \lambda^-(v),
\end{equation}
and for $v<0$,
\begin{equation}\label{G(v)-}
G(v)  =  \frac{\omega_a+\omega_b}{2}+ \frac{\sqrt{\Omega}}{2v}
\left(Y^+(v) - \frac{1}{Y^+(v)} \right) - v \lambda^+(v),
\end{equation}
where
\begin{eqnarray}
Y^\pm(v) & = & \frac{1}{2} \left( Z(v) \pm
\sqrt{Z(v)^2-4} \right), \nonumber \\
 \lambda^\pm(v) & = & - \frac{\Psi}{2} + \frac{1}{2} \ln \left( \frac{  Z(v)
\pm \sqrt{Z(v)^2-4}  }{2} \right), \label{Y and lambda}
\end{eqnarray}
and
\begin{equation}\label{Z}
Z(v)=\frac{v^2}{\sqrt{\Omega}} + \left( \frac{v^4}{\Omega} + 4 +
\frac{v^2 \Sigma^2}{\Omega} \right)^{1/2},
\end{equation}
with the following parameters
\begin{eqnarray}
\Omega & = & 4 \Wr_a \Wr_b \Wl_a \Wl_b, \label{parameters1} \\
\Sigma^2 & = & \left( \omega_a + \omega_b \right)^2 - 4 \left(
\Wl_a \Wl_b+ \Wr_a \Wr_b \right),
 \label{parameters2}
\end{eqnarray}
and $\Psi$ is the effective potential defined in
Eq.~\ref{Effective potential}.
\begin{figure}
\includegraphics[height=1.9in]{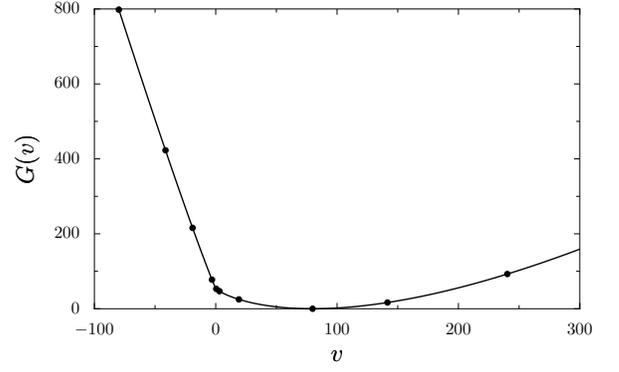}
\caption{Large deviation function $G(v)$, the solid line is the
exact expression using Eqs.~\ref{G(v)+}-\ref{G(v)-} and the points
are the numerical evaluation of the Legendre transform using
Eq.~\ref{explicit LD}. For the values of the rates used here,
 the average velocity, as  given by  Eq.~\ref{explicit v}, is $\bar{v}\simeq 80$:
   thus,  the system is far from equilibrium. Note
that  $G(v)$ is  minimum  at  $\bar{v}$ and  that  $G(\bar{v})=0$.}
\label{fig:LD}
\end{figure}
As  shown in fig.~\ref{fig:LD}, the function $G(v)$ has a single
minimum at the average velocity $v=\bar{v}$, which was defined in
Eq.~\ref{explicit v}, and at this point $G(v=\bar{v})=0$, which
can be deduced from Eq.~\ref{explicit LD}. Remarkably, although
$G(v)$ is a complicated non-linear function of $v$, the difference
$G(v)-G(-v)$ is a simple linear function of $v$ as required by the
Fluctuation Theorem. Using  the expressions~(\ref{explicit LD})
and~(\ref{G(v)+}), it is straightforward to verify  that
\begin{equation}\label{GC for LD}
G(v)-G(-v)=  \Psi v,
\end{equation}
which in turns implies that the ratio of the probabilities to
observe $v$ or $-v$ for large $t$ must obey:
\begin{equation}\label{GC for P}
\frac{\mathcal{P}(\frac{n}{t}=v)}{\mathcal{P}(\frac{n}{t}=-v)}=e^{-
\Psi vt}.
\end{equation}
From Eq.~\ref{GC for LD}, and the fact that $G(v)$ and
$\vartheta(\lambda)$ are related by a Legendre transform, we
obtain a third formulation of the Fluctuation Theorem:
\begin{equation}\label{GC for theta}
\vartheta(\lambda)= \vartheta( -\Psi- \lambda).
\end{equation}
Near equilibrium, the large deviation function is well
approximated by a half parabola when both $v$ when $\bar{v}$ are either
 positive or negative.  For  $\bar{v}>0$, this part of the large
deviation function becomes flatter and flatter, when going away
from equilibrium ({\it i.e.},  for increasing entropy production). As
a result, the remaining part of the large deviation function for
$v<0$ must be linear $G(v<0) \simeq -\Psi v$, so that Eq.~\ref{GC
for LD} is obeyed. This linear part for $v<0$ and the half
parabola for $v>0$ can be seen in fig.~\ref{fig:LD}.

It is interesting to note the central role played by the quantity
$\Psi$ defined initially as an effective potential, and which now
enters the three formulations of the Fluctuation Theorem in
Eqs.~\ref{GC for LD}-\ref{GC for theta}. Note that these equations
were obtained for arbitrary forms  of the rates
$\Wr_a,\Wr_b,\Wl_a$ and $\Wl_b$. If we make a specific choice for
these rates as in
 Eq.~\ref{rates}, with no chemistry {\it i.e.},  for $\Delta
\mu=0$, we recover $\Psi=-f$, and then Eq.~\ref{GC for theta}
reduces to $\vartheta(f-\lambda)=\vartheta(\lambda)$, which is
indeed compatible with Eq.~\ref{FT} when there is no chemical
variable and no dependance on the rates on chemistry. If the rates
are those of Eq.~\ref{rates} for $\Delta \mu \neq 0$, we obtain
using Eq.~\ref{Effective potential2}, the expression
\begin{equation}\label{GC for P2}
\frac{\mathcal{P}(\frac{n}{t}=v)}{\mathcal{P}(\frac{n}{t}=-v)}=e^{
(f-f_{st}(\Delta \mu)) vt}, \,\,\,\, (\rm{for} \, t \rightarrow
\infty)
\end{equation}
with the stalling force defined in Eq.~\ref{Stalling force}, and
related to $\Psi$ by $\Psi=-f+f_{st}(\Delta \mu)$.

Note that an Einstein relation can be obtained near stalling, by
performing a Taylor expansion of the r.h.s of Eq.~\ref{GC for
theta} with respect to $-\Psi$, in way similar to what was done in
Eqs.~\ref{Taylor1}-\ref{Taylor2} for the derivation of the
Einstein and Onsager relations. This procedure means that for $f
\simeq f_{st}(\Delta \mu)$,
\begin{equation}\label{Einstein rel stalling}
\bar{v} \simeq \frac{1}{2} \left(f - f_{st}(\Delta \mu) \right)
\left. \frac{\partial^2 \vartheta}{\partial^2 \lambda} \right|_0,
\end{equation}
which shows that near the stalling force, the Einstein relation
holds in this description where the chemical variable is absent
\cite{prl}.

\subsection{Discussion}
Note that Eq.~\ref{GC for P2} puts a constraint on the ratio of
the probabilities of observing a velocity $v$ to the probability
of observing a velocity $-v$. These velocities should be estimated
from the ratio $n/t$ based on an observation of the motor running
a distance $n$ (or a distance $-n$), after a time $t$. This
relation has been proven here in the limit of long time $t$, but
we expect that such a relation will also hold at finite time $t$
under some conditions, as suggested by our derivation of the
transient FT of Eq.~\ref{FT finite time}. Such a relation at
finite time was also investigated in Ref.~\cite{gaspard}.

Single molecule experiments on kinesin in which backward steps
were studied were performed in Refs.~\cite{nishiyama,carter}. In
particular it was shown in these references that ATP binding was
necessary for backward steps, and that the ratio of the overall
probability of making one forward step (whatever the time) to the
overall probability of making one backward step (whatever the
time) is an exponential function of the load, which approaches one
near the stalling force. It is important to point out that this
ratio which was measured experimentally is not the same quantity
as the left hand side of Eq.~\ref{GC for P2} although both
quantities should be related. In view of this, Eq.~\ref{GC for P2}
should be considered as a prediction for the behavior of single
motors like kinesin, which to our knowledge has not yet been
tested experimentally. This suggests that it would be very
interesting to probe Eq.~\ref{GC for P2} experimentally, by trying
to compute directly the distributions $\mathcal{P}(\frac{n}{t}=\pm
v)$ for various times. At the same time, it would be also useful
to study more extensively the behavior of motors of various types
near stalling as function of the ATP concentration. No notable
difference could be measured in the stalling force at an ATP
concentration $1m$M or $10 \mu$M in the experiments of
\cite{carter} on kinesin, although in principle according to
general grounds \cite{lipowsky,liepelt} one should expect that the
behavior of motors near stalling (and in particular the stalling
force of the motor itself) should depend on the ATP concentration
and more generally on the details of the chemical cycle of ATP
hydrolysis.

\subsection{Other forms of FT relations}
We have seen  that the form of the FT relation depends on the
state variables of the system, or in other words, it depends
 on the level of  coarse-graining of
  the description. We consider in this paper the following levels
of description:
\begin{itemize}
    \item (I): The mechanical displacement  $n$ is the only state
    variable.
    \item (II): The chemical variable $y$ is the only state
    variable.
    \item (III): Both variables $n$ and $y$ are taken into
    account.
\end{itemize}

For case (I), the dynamics is described by the simplified master
equation, Eq.~\ref{1 var Master eq} and Eqs.~(\ref{GC for
LD}-\ref{GC for theta}) are the appropriate forms of FT.

  For case (II), we find that FT can be written in the
following forms (we shall omit the large deviation function form
of FT  for cases (II) and (III)):
\begin{equation}\label{GC for P chimie}
\frac{\mathcal{P}(\frac{y}{t}=r)}{\mathcal{P}(\frac{y}{t}=-r)}=e^{-
\chi r t},
\end{equation}
which also holds generally for $t \rightarrow \infty$ and
\begin{equation}\label{GC for theta chimie}
\vartheta(\gamma)= \vartheta( -\chi- \gamma).
\end{equation}
In the above two equations, $-\chi$ can be interpreted as the
affinity \cite{deDonder} associated with a chemical cycle (see the
representation of the chemical cycle in
fig.~\ref{fig:representation}, and next section for a discussion
on the notion of affinity). We find that $\chi$  is given by
\begin{equation}\label{Effective potential_chimie} \chi=
 \ln \left( \frac{\omega_b^{-1} \omega_a^0}{\omega_a^1
\omega_b^0} \right),
\end{equation}
where $\omega_a^l=\Wr_a^l+\Wl_a^l$ and
$\omega_b^l=\Wr_b^l+\Wl_b^l$ for $l=-1,0$ or $1$. The physical
interpretation of $\chi$  can be clarified by using a method
similar to \cite{nelson}:  after integrating out the position
variable $n$ from  Eq.~\ref{Master eq},  and decimating over the
odd or even sites, one can derive  an effective evolution equation
for the occupation probabilities of  the remaining sites. In this
equation, $\chi$ plays the role of a effective potential for the
chemical variable.   When the rates of Eq.~\ref{rates} are used,
we find that this quantity is given by
\begin{equation} \label{eq:affinitepurechimie}
 \chi = -\Delta\mu + \Delta\mu_{st}(f) \, .
\end{equation}
As expected, the conditions for which $\chi$ vanishes are the same
as those for which  the chemical current $r$,
  given  in Eq.~\ref{explicit r},  vanishes.

  For case (III), the FT can be written as follows
\begin{equation}\label{GC for P meca-chimie}
\frac{\mathcal{P}(\frac{n}{t}=v,\frac{y}{t}=r)}{\mathcal{P}(\frac{n}{t}=
-v,\frac{y}{t}=-r)}=e^{-(\tilde{\Psi} v +\tilde{\chi} r)t},
\end{equation}
for $t \rightarrow \infty$ and
\begin{equation}\label{GC for theta meca-chimie}
\vartheta(\lambda,\gamma)= \vartheta(-\tilde{\Psi} - \lambda,
-\tilde{\chi}- \gamma).
\end{equation}
 Here,    the affinities associated with the mechanical and chemical
variables  are given,  respectively,  by
  $-\tilde{\Psi}$ and $-\tilde{\chi}$.
 Note that these quantities are in
general  not the same as the ones calculated above in
cases (I) and (II) ({\it i.e.}, $\tilde{\Psi} \neq \Psi$
 and  $ \tilde{\chi} \neq  \chi$).
 When the rates of
Eq.~\ref{rates} are used, $\tilde{\Psi}=-f$ and
$\tilde{\chi}=-\Delta \mu$, so that Eq.~\ref{FT} is recovered from
Eq.~\ref{GC for theta meca-chimie}.

\section{Fluctuation Theorem and Entropy Production}
 In this section, we discuss the connections
  between the Fluctuation Theorem described in the last section and
 the entropy production~\cite{lebowitz}. In particular, we show
 by  an explicit calculation, that the
 parameters $\Psi$, $\chi$, $f$ and $\Delta\mu$ that appear
 in the symmetry relations Eqs.~\ref{GC for
LD}-\ref{GC for P2} are identical to the
  affinities  associated with the various
 macroscopic currents (mechanical and chemical)
 flowing in the system~\cite{gaspard}.
 Affinities, introduced  a long time ago
 in  chemical thermodynamics  \cite{deDonder},
   represent  intrinsic quantities
  that depend only  on  the microscopic transition rates
 of the system.   Thus,
 the fact that these  quantities also  appear  in the Fluctuations
 Theorems,  valid far from equilibrium,  shows   a remarkable
 connection between classical  thermodynamics and
 non-equilibrium statistical mechanics.

We shall first discuss the simplified model, in which the chemical
variable $y$ is not taken into account in the description as a
state variable (case (I)). The mechanical entropy $S_{\rm M}(t)$
then only contains contribution from the disorder in the
distribution of the mechanical
  variable $n$ and is defined as
\begin{equation} \label{def:entropy}
 S_{\rm M}(t)=-\sum _{i=a,b}\sum _n P_i(n,t) \ln P_i(n,t)
\end{equation}
in units where $k_B=1$. Using the master equation,  Eq.~\ref{1 var
Master eq}, one can calculate the variation of $S_M(t)$ with time:
\begin{eqnarray}
 \frac{dS_{\rm M}}{dt} =  {\hskip 6cm} \\ \sum_{i\neq j} \sum _n
       \left( \overrightarrow{\omega}_i  P_i(n,t) -
     \overleftarrow{\omega}_j   P_j(n+1,t) \right)
  \ln\frac{P_i(n,t)}{P_j(n+1,t)} \nonumber \, ,
\end{eqnarray}
 where $i,j$ take the two possible values $a$ and $b$ but are different from each other.
 By transforming the last term in this equation  as follows
 $$ \ln\frac{P_i(n,t)}{P_j(n+1,t)} =
  \ln\frac{\overrightarrow{\omega}_i  P_i(n,t))}
 {\overleftarrow{\omega}_j P_j(n+1,t)}
    -  \ln\frac{\overrightarrow{\omega}_i}{\overleftarrow{\omega}_j} \, , $$
  the time derivative of the entropy can be  rewritten as
  the difference of  an entropy production (which is always positive)
 and an entropy flux. Since we
are interested in a stationary state where $dS/dt=0$, both
contributions must be equal. From such a calculation, one finds
that the entropy production and  the entropy flux
 in the long time limit are given by
\begin{equation}\label{EntropyProd}
\Pi_{\rm M}= \frac{\Wr_a \Wr_b - \Wl_a
\Wl_b}{\Wr_a+\Wr_b+\Wl_a+\Wl_b}
 \ln \left( \frac{\Wr_a \Wr_b}{\Wl_a \Wl_b} \right)  =    -\Psi \bar{v},
\end{equation}
where $\Psi$ is the effective potential defined in
Eq.~\ref{Effective potential}, and $\bar{v}$ is defined in
Eq.~\ref{explicit v}. According to the general definition
\cite{gaspard}, we deduce from Eq.~\ref{EntropyProd} that the
mechanical affinity of the displacement variable is $-\Psi$.

It is interesting to recall that
 this result can also be derived in a different way~:
   in~\cite{lebowitz}, it was
proven that the entropy flux  can be calculated by using
  a fluctuating quantity $W(t)$, called
 the action functional,  which can be
seen as a local measure of the lack of detailed balance on a given
path at time $t$.
 The matrix ${\cal N}(\nu)$  that describes
 the  evolution of the  generating
function of   $\langle \exp(-\nu W(t)) \rangle$   is given by
\begin{eqnarray} {\cal N}= \left [
 \begin{array}{cc}
  -\overrightarrow{\omega_a}-\overleftarrow{\omega_a} &
  \, \overleftarrow{\omega_a}^{(1-\nu)} \overrightarrow{\omega_b}^\nu+
   \overleftarrow{\omega_b}^{\nu} \overrightarrow{\omega_a}^{(1-\nu)}\\
   \overleftarrow{\omega_b}^{(1-\nu)} \overrightarrow{\omega_a}^\nu+
   \overleftarrow{\omega_a}^{\nu} \overrightarrow{\omega_b}^{(1-\nu)}
   & -\overleftarrow{\omega_b}-\overrightarrow{\omega_b} \\ \nonumber
\end{array} \right ].  \\ \nonumber
\end{eqnarray}
 This  matrix   is  obtained by deforming
 the original Markov matrix by a parameter $\nu$.
 We emphasize that this deformation is not the same  as that
  used in Eq.~\ref{Matrice M(lambda)}
  to calculate the large deviation of the currents.
 The time derivative of  $W(t)$ is precisely
 the entropy flux. Therefore, we have, in agreement
  with Eq.~\ref{EntropyProd},
\begin{equation}\label{Entropy from N}
\Pi_{\rm M}=-\frac{\partial q}{\partial \nu}(0)=-\bar{v} \Psi,
\end{equation}
 where $q(\nu)$ is the  largest eigenvalue  of  ${\cal N}(\nu)$.
 Since the  matrix ${\cal N}$  has the property
  ${\cal N}^{\dag}(\nu) = {\cal N}(1-\nu)$,
   its  largest eigenvalue $q(\nu)$ satisfies a Fluctuation Theorem
\begin{equation}\label{FT for N}
q(1-\nu)=q(\nu) \, .
\end{equation}
 Note that Eq.~\ref{EntropyProd} is valid for arbitrary transition
 rates;   if we
make a specific choice for the rates such as that of
Eq.~\ref{rates}, we find that $\Pi=(-f+f_{st}(\Delta \mu))
\bar{v}$ when $\Delta \mu \neq 0$. When $\Delta \mu=0$, we recover
$\Pi=-f \bar{v}$, a well-known result \cite{parmeggiani}, but  our
  more general expression for
$\Delta \mu \neq 0$  shows explicitly the dependence of the
entropy production on a measurable quantity $\Delta \mu$ and its
connection to the FT which we saw in Eq.~\ref{GC for P2}.

 If we now use  a description of the model where only
  the chemical variable $y$ is taken into account (case (II))
 and the total displacement $n$ is integrated out, we can define
 a `chemical entropy' $S_{\rm C}(t)$ as follows
\begin{equation} \label{def:chementropy}
 S_{\rm C}(t)=-\sum _{i=a,b}\sum _y P_i(y,t) \ln P_i(y,t) \, .
\end{equation}
 Calculations similar to the ones described above allow us
 to derive  the purely chemical entropy production in the stationary state:
\begin{equation}\label{ChemEntropyProd}
\Pi_{\rm C}=  \frac{\omega_a^1 \omega_b^0 - \omega_a^0
\omega_b^{-1}}{\omega_a + \omega_b} \ln \left( \frac{\omega_a^1
\omega_b^0}{\omega_b^{-1} \omega_a^0} \right)  = -r \chi \, ;
\end{equation}
  the chemical current $r$ and the chemical affinity
 $\chi$ were defined in Eq.~\ref{explicit r} and
 Eq.~\ref{Effective potential_chimie} respectively.
 We also note that the entropy production in Eq.~\ref{ChemEntropyProd}
  can  be calculated  using an  action functional whose generating function
 is the largest eigenvalue of the Markov matrix
   suitably  deformed  \cite{lebowitz}.

Finally, we can  use the  complete description of
 Eq.~\ref{Master eq}, in which both the displacement  $n$
 and   the chemical variable $y$ are  taken into account (case (III)). In this case,
the entropy is given by
\begin{equation} \label{def:totalentropy}
 S(t)=-\sum _{i=a,b}\sum _n\sum _y P_i(n,y,t) \ln P_i(n,y,t) \, .
\end{equation}
 Again, if we make the
specific choice for the rates of Eq.~\ref{rates}, we find that the
following well-known result
 \cite{parmeggiani}  is recovered
 for the entropy production:
\begin{equation}\label{Gen Entropy}
\Pi= f \bar{v} + r \Delta \mu \, .
\end{equation}
 This relation
  makes explicit the fact that $f$ is the affinity of  the
  mechanical  position
variable with the current $\bar{v}$,  and  that
 $\Delta \mu$  is the affinity
of the chemical variable with the current $r$.
  We note that these affinities are different from those
 found above in the purely mechanical and in
 the purely chemical models, which correspond respectively
to   Eq.~\ref{Effective potential}  and to
Eq.~\ref{eq:affinitepurechimie}.
  The fact that the expression
 of the  entropy (and hence that of the affinity)
  strongly  depends on the level of coarse-graining used in a
 given description should not come as a surprise.
The two affinities $f$  and  $\Delta \mu$ appear in the
Gallavotti-Cohen relation Eq.~\ref{FT}. This suggests that  one
should be able
  to construct an effective potential describing the evolution
of the motor in a 2 dimensional phase space of $n$ and $y$, and
that this potential should be equivalent to the potential of mean
force discussed in Ref.~\cite{bustamente}.

  We have seen here  that  the FT for
the currents and the FT for the entropy are closely related; this
 fact is true  for a large class of models as explained in
  Ref.~\cite{lebowitz}.  However, although
 the FT for the entropy holds
generally for any markovian dynamics as shown in
Ref.~\cite{lebowitz},  a  FT for the currents exists
 only if the dynamics can be decomposed into cycles  with  well
defined affinities \cite{gaspard}.
 This is why the periodicity of the motion of the
motor along track and of the evolution of the chemical variable
was a crucial  assumption in  our derivation
 of  the FT for the currents but was not used when
deriving the FT for the entropy.

\section{Conclusion}
We have studied a discrete stochastic model of a molecular motor,
which is a minimal ratchet model. We made contact in this paper
between various formulations of FT. Through a detailed analysis of
a simple model, we have brought out some physical implications of
FT for molecular motors in general and for kinesin in particular.
 One important message is that FT puts constraints on
the operation of a molecular motor or nano-machines far from
equilibrium. Further experimental work and theoretical modelling
is necessary to check more precisely the implications of FT for
molecular motors. For instance, it would be interesting to study a
molecular motor in which both the velocity and the average ATP
consumption rate could be measured simultaneously, or if this is
too difficult study more extensively the behavior of motors near
the stalling force as function of ATP concentration. This would
allow a study of the violations of the Fluctuation-Dissipation at
the level of a single motor, which would lead to much deeper
insights into the Mechano-transduction mechanism of molecular
motors.

Due to the broad applicability of the ratchet concept in
biological systems, we believe that the results of this paper
should be of general applicability: the model could describe
processive molecular motors of various types, nano-machines like
enzymes performing chemical cycles or polymers which are
translocated through a pore under the action of a force (for
instance the force created by an electric field  applied to a charged
polymer). More generally, we hope that the present work
illustrates the usefulness of statistical physics of
non-equilibrium systems for the understanding of active systems,
and in particular biological systems.

We acknowledge stimulating discussions with A.\ Ajdari, M.\
Schindler, J.F.\ Joanny, F. J\"{u}licher, and J.\ Prost. We thank
C. Schmidt for pointing out to us Ref.~\cite{carter}. A.W.C.L.
acknowledges support from the ESPCI (Chaire Joliot) and from NSF
Grant No. DMR-0701610 (for A.W.C.L.). D.L. acknowledges support
from
the Indo-French Center for grant 3504-2. \\

\appendix
\section{Calculation of the large deviation function G(v)}
To obtain an explicit form for $G(v)$, we  take  the square of
Eq.~\ref{formeparamLD}, \bb \label{Derivative U2} \left(
\frac{\partial \vartheta}{\partial \lambda}[\lambda=\lambda^*]
\right)^2  = v^2
 .  \en
Using Eq.~\ref{explicittheta} the derivative on the left hand side
can be written as
\begin{equation}\label{Derivative}
\frac{\partial \vartheta}{\partial \lambda}= \frac{1}{4}
\frac{U'(\lambda)}{\sqrt{U(\lambda)}},
\end{equation}
with $U(\lambda)=(\omega_a - \omega_b )^2
  +  4 ( \Wl_b e^\lambda + \Wr_b
e^{-\lambda} )  (  \Wl_a e^\lambda + \Wr_a e^{-\lambda} )$. After
performing the change of variable \bb \label{change of variable}
Y=e^{2\lambda} \sqrt{\frac{\Wl_a \Wl_b}{\Wr_a \Wr_b}}, \en and
using the parameters $\Omega$ and $\Sigma$ introduced in
Eqs.~\ref{parameters1}-\ref{parameters2}, we can write
\begin{equation}\label{Up}
U'(\lambda)=4 \sqrt{\Omega} \left( Y- \frac{1}{Y} \right),
\end{equation}
and Eq.~\ref{Derivative U2} becomes
 \bb 16 \Omega \left(
Y-\frac{1}{Y} \right)^2 = 16 v^2 [ \Sigma^2 + 2 \sqrt{\Omega}
\left( Y+\frac{1}{Y} \right) ]. \en
 We deduce   that  $Z=Y+1/Y$  satisfies   \bb Z^2-
\frac{2v^2}{\sqrt{\Omega}} Z - 4 - \frac{v^2 \Sigma^2}{\Omega}=0.
\en There are two solutions to this equation but since $Z>0$, only
the positive solution must be retained which is Eq.~\ref{Z}. To obtain
$Y$ in terms of $Z(v)$, one must solve another second order
equation $Y^2-Z Y + 1=0$. This  equation has  two positive
acceptable solutions, which are the two solutions $Y^\pm(v)$ of
Eq.~\ref{Y and lambda}.  We have
 $Y^{+}(v)> Y^{-}(v) = 1/ Y^{+}(v).$
  Using  Eq.~\ref{formeparamLD} and Eq.~\ref{Derivative},
  we see that  $Y^{+}(v)$ corresponds to
 $v<0$. Similarly,  $ Y^{-}(v)$ corresponds to $v<0$. Once the
relation $Y=Y(v)$ is determined, it  is easily inverted
using Eq.~\ref{change of variable} to yield $\lambda^*(v)$, which
is precisely $\lambda^\pm(v)$ in the second equation of Eq.~\ref{Y
and lambda}. The final expression of $G(v)$ is  obtained by
substituting  this result into Eq.~\ref{explicit LD}.

\bibliographystyle{apsrev}
\bibliography{onsager}

\begin{thebibliography}{39}
\expandafter\ifx\csname natexlab\endcsname\relax\def\natexlab#1{#1}\fi
\expandafter\ifx\csname bibnamefont\endcsname\relax
  \def\bibnamefont#1{#1}\fi
\expandafter\ifx\csname bibfnamefont\endcsname\relax
  \def\bibfnamefont#1{#1}\fi
\expandafter\ifx\csname citenamefont\endcsname\relax
  \def\citenamefont#1{#1}\fi
\expandafter\ifx\csname url\endcsname\relax
  \def\url#1{\texttt{#1}}\fi
\expandafter\ifx\csname urlprefix\endcsname\relax\def\urlprefix{URL }\fi
\providecommand{\bibinfo}[2]{#2}
\providecommand{\eprint}[2][]{\url{#2}}

\bibitem[{\citenamefont{Martin et~al.}(2001)\citenamefont{Martin, Hudspeth, and
  J{\"u}licher}}]{martin}
\bibinfo{author}{\bibfnamefont{P.}~\bibnamefont{Martin}},
  \bibinfo{author}{\bibfnamefont{A.~J.} \bibnamefont{Hudspeth}},
  \bibnamefont{and}
  \bibinfo{author}{\bibfnamefont{F.}~\bibnamefont{J{\"u}licher}},
  \bibinfo{journal}{Proc. Natl. Acad. Sci.} \textbf{\bibinfo{volume}{98}},
  \bibinfo{pages}{14380} (\bibinfo{year}{2001}).

\bibitem[{\citenamefont{Camalet et~al.}(2000)\citenamefont{Camalet, Duke,
  J{\"u}licher, and Prost}}]{duke}
\bibinfo{author}{\bibfnamefont{S.}~\bibnamefont{Camalet}},
  \bibinfo{author}{\bibfnamefont{T.}~\bibnamefont{Duke}},
  \bibinfo{author}{\bibfnamefont{F.}~\bibnamefont{J{\"u}licher}},
  \bibnamefont{and} \bibinfo{author}{\bibfnamefont{J.}~\bibnamefont{Prost}},
  \bibinfo{journal}{Proc. Natl. Acad. Sci.} \textbf{\bibinfo{volume}{97}},
  \bibinfo{pages}{3183} (\bibinfo{year}{2000}).

\bibitem[{\citenamefont{Manneville et~al.}(2001)\citenamefont{Manneville,
  Bassereau, Ramaswamy, and Prost}}]{jb_PRE}
\bibinfo{author}{\bibfnamefont{J.~B.} \bibnamefont{Manneville}},
  \bibinfo{author}{\bibfnamefont{P.}~\bibnamefont{Bassereau}},
  \bibinfo{author}{\bibfnamefont{S.}~\bibnamefont{Ramaswamy}},
  \bibnamefont{and} \bibinfo{author}{\bibfnamefont{J.}~\bibnamefont{Prost}},
  \bibinfo{journal}{Phys. Rev. E} \textbf{\bibinfo{volume}{64}},
  \bibinfo{pages}{021908} (\bibinfo{year}{2001}).

\bibitem[{\citenamefont{Kruse et~al.}(2003)\citenamefont{Kruse, Joanny,
  J{\"u}licher, Prost, and Sekimoto}}]{joanny}
\bibinfo{author}{\bibfnamefont{K.}~\bibnamefont{Kruse}},
  \bibinfo{author}{\bibfnamefont{J.}~\bibnamefont{Joanny}},
  \bibinfo{author}{\bibfnamefont{F.}~\bibnamefont{J{\"u}licher}},
  \bibinfo{author}{\bibfnamefont{J.}~\bibnamefont{Prost}}, \bibnamefont{and}
  \bibinfo{author}{\bibfnamefont{K.}~\bibnamefont{Sekimoto}},
  \bibinfo{journal}{Eur. Phys. J. E} \textbf{\bibinfo{volume}{91}},
  \bibinfo{pages}{198101} (\bibinfo{year}{2003}).

\bibitem[{\citenamefont{Mizuno et~al.}(2007)\citenamefont{Mizuno, Tarding,
  Schmidt, and MacKintosh}}]{fred}
\bibinfo{author}{\bibfnamefont{D.}~\bibnamefont{Mizuno}},
  \bibinfo{author}{\bibfnamefont{C.}~\bibnamefont{Tarding}},
  \bibinfo{author}{\bibfnamefont{C.~F.} \bibnamefont{Schmidt}},
  \bibnamefont{and} \bibinfo{author}{\bibfnamefont{F.~C.}
  \bibnamefont{MacKintosh}}, \bibinfo{journal}{Science}
  \textbf{\bibinfo{volume}{315}}, \bibinfo{pages}{370} (\bibinfo{year}{2007}).

\bibitem[{\citenamefont{Mayor and Rao}(2004)}]{madan_mayor}
\bibinfo{author}{\bibfnamefont{S.}~\bibnamefont{Mayor}} \bibnamefont{and}
  \bibinfo{author}{\bibfnamefont{M.}~\bibnamefont{Rao}},
  \bibinfo{journal}{Traffic} \textbf{\bibinfo{volume}{5}}, \bibinfo{pages}{231}
  (\bibinfo{year}{2004}).

\bibitem[{\citenamefont{Lau et~al.}(2003)\citenamefont{Lau, Hoffman, Davies,
  Crocker, and Lubensky}}]{andy}
\bibinfo{author}{\bibfnamefont{A.~W.~C.} \bibnamefont{Lau}},
  \bibinfo{author}{\bibfnamefont{B.~D.} \bibnamefont{Hoffman}},
  \bibinfo{author}{\bibfnamefont{A.}~\bibnamefont{Davies}},
  \bibinfo{author}{\bibfnamefont{J.~C.} \bibnamefont{Crocker}},
  \bibnamefont{and} \bibinfo{author}{\bibfnamefont{T.~C.}
  \bibnamefont{Lubensky}}, \bibinfo{journal}{Phys. Rev. Lett.}
  \textbf{\bibinfo{volume}{91}}, \bibinfo{pages}{198101}
  (\bibinfo{year}{2003}).

\bibitem[{\citenamefont{Howard}(2001)}]{howard}
\bibinfo{author}{\bibfnamefont{J.}~\bibnamefont{Howard}},
  \emph{\bibinfo{title}{Mechanics of Motor Proteins and the Cytoskeleton}}
  (\bibinfo{publisher}{Sinauer Associates}, \bibinfo{address}{Sunderland, MA},
  \bibinfo{year}{2001}).

\bibitem[{\citenamefont{Asbury}(2005)}]{motorreview}
\bibinfo{author}{\bibfnamefont{C.}~\bibnamefont{Asbury}},
  \bibinfo{journal}{Curr. Opin. Cell Biol.} \textbf{\bibinfo{volume}{17}},
  \bibinfo{pages}{89} (\bibinfo{year}{2005}).

\bibitem[{\citenamefont{J{\"u}licher et~al.}(1997)\citenamefont{J{\"u}licher,
  Ajdari, and Prost}}]{armand1}
\bibinfo{author}{\bibfnamefont{F.}~\bibnamefont{J{\"u}licher}},
  \bibinfo{author}{\bibfnamefont{A.}~\bibnamefont{Ajdari}}, \bibnamefont{and}
  \bibinfo{author}{\bibfnamefont{J.}~\bibnamefont{Prost}},
  \bibinfo{journal}{Rev. Mod. Phys.} \textbf{\bibinfo{volume}{69}},
  \bibinfo{pages}{1269} (\bibinfo{year}{1997}).

\bibitem[{\citenamefont{Parmeggiani et~al.}(1999)}]{parmeggiani}
\bibinfo{author}{\bibfnamefont{A.}~\bibnamefont{Parmeggiani}}
  \bibnamefont{et~al.}, \bibinfo{journal}{Phys. Rev. E}
  \textbf{\bibinfo{volume}{60}}, \bibinfo{pages}{2127} (\bibinfo{year}{1999}).

\bibitem[{\citenamefont{Kolomeisky and Widom}(1998)}]{widom}
\bibinfo{author}{\bibfnamefont{A.}~\bibnamefont{Kolomeisky}} \bibnamefont{and}
  \bibinfo{author}{\bibfnamefont{B.}~\bibnamefont{Widom}}, \bibinfo{journal}{J.
  Stat. Phys.} \textbf{\bibinfo{volume}{93}}, \bibinfo{pages}{633}
  (\bibinfo{year}{1998}).

\bibitem[{\citenamefont{Fisher and Kolomeisky}(1999)}]{kolomeisky}
\bibinfo{author}{\bibfnamefont{M.}~\bibnamefont{Fisher}} \bibnamefont{and}
  \bibinfo{author}{\bibfnamefont{A.}~\bibnamefont{Kolomeisky}},
  \bibinfo{journal}{Proc. Natl. Acad. Sci.} \textbf{\bibinfo{volume}{96}},
  \bibinfo{pages}{6597} (\bibinfo{year}{1999}), \bibinfo{note}{{\it ibid.},
  {\bf 98}, 7748 (2001).}

\bibitem[{\citenamefont{Kafri et~al.}(2004)}]{nelson}
\bibinfo{author}{\bibfnamefont{Y.}~\bibnamefont{Kafri}} \bibnamefont{et~al.},
  \bibinfo{journal}{Biophys. J.} \textbf{\bibinfo{volume}{86}},
  \bibinfo{pages}{3373} (\bibinfo{year}{2004}).

\bibitem[{\citenamefont{Jarzynski and Mazonka}(1999)}]{mazonka}
\bibinfo{author}{\bibfnamefont{C.}~\bibnamefont{Jarzynski}} \bibnamefont{and}
  \bibinfo{author}{\bibfnamefont{O.}~\bibnamefont{Mazonka}},
  \bibinfo{journal}{Phys. Rev. E} \textbf{\bibinfo{volume}{59}},
  \bibinfo{pages}{6448} (\bibinfo{year}{1999}).

\bibitem[{\citenamefont{Lipowsky}(2000)}]{lipowsky}
\bibinfo{author}{\bibfnamefont{R.}~\bibnamefont{Lipowsky}},
  \bibinfo{journal}{Phys. Rev. Lett.} \textbf{\bibinfo{volume}{85}},
  \bibinfo{pages}{4401} (\bibinfo{year}{2000}), \bibinfo{note}{{G}. Lattanzi
  and A. Maritan, Phys. Rev. Lett. {\bf 86} 1134 (2001).}

\bibitem[{\citenamefont{Schnitzer and Block}(1997)}]{block}
\bibinfo{author}{\bibfnamefont{M.}~\bibnamefont{Schnitzer}} \bibnamefont{and}
  \bibinfo{author}{\bibfnamefont{S.}~\bibnamefont{Block}},
  \bibinfo{journal}{Nature} \textbf{\bibinfo{volume}{388}},
  \bibinfo{pages}{386} (\bibinfo{year}{1997}), \bibinfo{note}{{K}. Visscher
  {\em et al.}, Nature {\bf 400}, 184 (1999)}.

\bibitem[{\citenamefont{Coppin et~al.}(1997)}]{coppin}
\bibinfo{author}{\bibfnamefont{C.}~\bibnamefont{Coppin}} \bibnamefont{et~al.},
  \bibinfo{journal}{Proc. Natl. Acad. Sci.} \textbf{\bibinfo{volume}{94}},
  \bibinfo{pages}{8539} (\bibinfo{year}{1997}), \bibinfo{note}{{M}. Nishiyama
  {\em et al.}, Nat. Cell Biol., {\bf 4}, 790 (2002); C.L. Asbury {\em et al.},
  Science {\bf 302}, 2130 (2003).}

\bibitem[{\citenamefont{Carter and Cross}(2005)}]{carter}
\bibinfo{author}{\bibfnamefont{N.}~\bibnamefont{Carter}} \bibnamefont{and}
  \bibinfo{author}{\bibfnamefont{R.}~\bibnamefont{Cross}},
  \bibinfo{journal}{Nature} \textbf{\bibinfo{volume}{435}},
  \bibinfo{pages}{308} (\bibinfo{year}{2005}).

\bibitem[{\citenamefont{Shaevitz et~al.}(2005)}]{schnitzer}
\bibinfo{author}{\bibfnamefont{J.}~\bibnamefont{Shaevitz}}
  \bibnamefont{et~al.}, \bibinfo{journal}{Biophys. J.}
  \textbf{\bibinfo{volume}{89}}, \bibinfo{pages}{2277} (\bibinfo{year}{2005}).

\bibitem[{\citenamefont{Evans et~al.}(1993)\citenamefont{Evans, Cohen, and
  Morriss}}]{FT}
\bibinfo{author}{\bibfnamefont{D.}~\bibnamefont{Evans}},
  \bibinfo{author}{\bibfnamefont{E.}~\bibnamefont{Cohen}}, \bibnamefont{and}
  \bibinfo{author}{\bibfnamefont{G.}~\bibnamefont{Morriss}},
  \bibinfo{journal}{Phys. Rev. Lett.} \textbf{\bibinfo{volume}{71}},
  \bibinfo{pages}{2401} (\bibinfo{year}{1993}), \bibinfo{note}{{G}. Gallavotti
  and E.G.D. Cohen, Phys. Rev. Lett., {\bf 74}, 2694 (1995); C. Jarzynski,
  Phys. Rev. Lett., {\bf 78}, 2690 (1997), J. Kurchan, J. Phys. A: Math. Gen.
  {\bf 31}, 3719 (1998); C. Maes, J. Stat. Phys. {\bf 95}, 367 (1999)}.

\bibitem[{\citenamefont{Gallavotti}(1996)}]{gallavotti}
\bibinfo{author}{\bibfnamefont{G.}~\bibnamefont{Gallavotti}},
  \bibinfo{journal}{Phys. Rev. Lett.} \textbf{\bibinfo{volume}{77}},
  \bibinfo{pages}{4334} (\bibinfo{year}{1996}).

\bibitem[{\citenamefont{Lebowitz and Spohn}(1999)}]{lebowitz}
\bibinfo{author}{\bibfnamefont{J.~L.} \bibnamefont{Lebowitz}} \bibnamefont{and}
  \bibinfo{author}{\bibfnamefont{H.}~\bibnamefont{Spohn}}, \bibinfo{journal}{J.
  Stat. Phys.} \textbf{\bibinfo{volume}{95}}, \bibinfo{pages}{333}
  (\bibinfo{year}{1999}).

\bibitem[{\citenamefont{Evans and Searles}(2002)}]{evans}
\bibinfo{author}{\bibfnamefont{D.}~\bibnamefont{Evans}} \bibnamefont{and}
  \bibinfo{author}{\bibfnamefont{D.}~\bibnamefont{Searles}},
  \bibinfo{journal}{Adv. Phys.} \textbf{\bibinfo{volume}{51}},
  \bibinfo{pages}{1529} (\bibinfo{year}{2002}), \bibinfo{note}{{\it ibid.}
  Phys. Rev. E {\bf 50}, 1645 (1994).}

\bibitem[{\citenamefont{Crooks}(2000)}]{crooks}
\bibinfo{author}{\bibfnamefont{G.~E.} \bibnamefont{Crooks}},
  \bibinfo{journal}{Phys. Rev. E} \textbf{\bibinfo{volume}{61}},
  \bibinfo{pages}{2361} (\bibinfo{year}{2000}).

\bibitem[{\citenamefont{Liphardt et~al.}(2002)}]{ritort}
\bibinfo{author}{\bibfnamefont{J.}~\bibnamefont{Liphardt}}
  \bibnamefont{et~al.}, \bibinfo{journal}{Science}
  \textbf{\bibinfo{volume}{296}}, \bibinfo{pages}{1832} (\bibinfo{year}{2002}),
  \bibinfo{note}{{D}. Collin {\em et al.}, Nature {\bf 437}, 231 (2005); V.
  Blickle {\em et al.}, Phys. Rev. Lett. {\bf 96}, 070603 (2006).}

\bibitem[{\citenamefont{Qian}(2005)}]{qian}
\bibinfo{author}{\bibfnamefont{H.}~\bibnamefont{Qian}}, \bibinfo{journal}{J.
  Phys.: Cond. Mat.} \textbf{\bibinfo{volume}{17}}, \bibinfo{pages}{S3783}
  (\bibinfo{year}{2005}).

\bibitem[{\citenamefont{Gaspard and Gerritsma}(2007)}]{gaspard}
\bibinfo{author}{\bibfnamefont{P.}~\bibnamefont{Gaspard}} \bibnamefont{and}
  \bibinfo{author}{\bibfnamefont{E.}~\bibnamefont{Gerritsma}},
  \bibinfo{journal}{J. Theo. Biol.} \textbf{\bibinfo{volume}{247}},
  \bibinfo{pages}{672} (\bibinfo{year}{2007}), \bibinfo{note}{{D}. Andrieux and
  P. Gaspard, Phys. Rev. E {\bf 74}, 011906 (2006); P. Gaspard, J. Chem. Phys.
  {\bf 120}, 8898 (2004).}

\bibitem[{\citenamefont{Seifert}(2005)}]{seifert}
\bibinfo{author}{\bibfnamefont{U.}~\bibnamefont{Seifert}},
  \bibinfo{journal}{Europhys. Lett.} \textbf{\bibinfo{volume}{70}},
  \bibinfo{pages}{36} (\bibinfo{year}{2005}).

\bibitem[{\citenamefont{Lau et~al.}(2007)\citenamefont{Lau, Lacoste, and
  Mallick}}]{prl}
\bibinfo{author}{\bibfnamefont{A.~W.~C.} \bibnamefont{Lau}},
  \bibinfo{author}{\bibfnamefont{D.}~\bibnamefont{Lacoste}}, \bibnamefont{and}
  \bibinfo{author}{\bibfnamefont{K.}~\bibnamefont{Mallick}},
  \bibinfo{journal}{Phys. Rev. Lett.} \textbf{\bibinfo{volume}{99}},
  \bibinfo{pages}{158102} (\bibinfo{year}{2007}).

\bibitem[{\citenamefont{Nishiyama et~al.}(2002)\citenamefont{Nishiyama,
  Higuchi, and Yanagida}}]{nishiyama}
\bibinfo{author}{\bibfnamefont{M.}~\bibnamefont{Nishiyama}},
  \bibinfo{author}{\bibfnamefont{H.}~\bibnamefont{Higuchi}}, \bibnamefont{and}
  \bibinfo{author}{\bibfnamefont{T.}~\bibnamefont{Yanagida}},
  \bibinfo{journal}{Nature Cell Biol.} \textbf{\bibinfo{volume}{4}},
  \bibinfo{pages}{790} (\bibinfo{year}{2002}).

\bibitem[{\citenamefont{Hill}(1987)}]{hill}
\bibinfo{author}{\bibfnamefont{T.~L.} \bibnamefont{Hill}},
  \emph{\bibinfo{title}{Linear Aggregation Theory in Cell Biology}}
  (\bibinfo{publisher}{Springer}, \bibinfo{address}{New York},
  \bibinfo{year}{1987}).

\bibitem[{\citenamefont{{De Donder}}(1927)}]{deDonder}
\bibinfo{author}{\bibfnamefont{T.}~\bibnamefont{{De Donder}}},
  \emph{\bibinfo{title}{L'Affinit{\'{e}}}}
  (\bibinfo{publisher}{Gauthiers-Villars}, \bibinfo{address}{Paris},
  \bibinfo{year}{1927}).

\bibitem[{not({\natexlab{a}})}]{note_on_delta_mu}
\bibinfo{note}{Note that this expression of $\Delta \mu$ requires that the
  solution be ideal. With the concentrations of ATP used in the experiments (at
  most a few mM), these solutions are sufficiently dilute that they can be
  considered ideal.}

\bibitem[{not({\natexlab{b}})}]{notation}
\bibinfo{note}{This change of sign is only due to a different convention for
  the orientation of the force: $F_e$ must be changed into $-F_e$ in the all
  the rates to recover the rates used in Ref.~\cite{nelson}. In our convention,
  a positive force corresponds to the direction of average motion of the
  motor.}

\bibitem[{\citenamefont{Derrida}(1983)}]{derrida}
\bibinfo{author}{\bibfnamefont{B.}~\bibnamefont{Derrida}}, \bibinfo{journal}{J.
  Stat. Phys.} \textbf{\bibinfo{volume}{31}}, \bibinfo{pages}{433}
  (\bibinfo{year}{1983}).

\bibitem[{\citenamefont{Sekimoto}(1997)}]{ken}
\bibinfo{author}{\bibfnamefont{K.}~\bibnamefont{Sekimoto}},
  \bibinfo{journal}{J. Phys. Soc. Jpn.} \textbf{\bibinfo{volume}{66}},
  \bibinfo{pages}{1234} (\bibinfo{year}{1997}), \bibinfo{note}{{\it ibid.}
  Phys. Rev. E, {\bf 76}, 060103(R) (2007).}

\bibitem[{\citenamefont{Parrondo and Espa{\~{n}}ol}(1996)}]{parrondo}
\bibinfo{author}{\bibfnamefont{J.~M.~R.} \bibnamefont{Parrondo}}
  \bibnamefont{and}
  \bibinfo{author}{\bibfnamefont{P.}~\bibnamefont{Espa{\~{n}}ol}},
  \bibinfo{journal}{Am. J. Phys.} \textbf{\bibinfo{volume}{64}},
  \bibinfo{pages}{1125} (\bibinfo{year}{1996}).

\bibitem[{\citenamefont{Keller and Bustamante}(2000)}]{bustamente}
\bibinfo{author}{\bibfnamefont{D.}~\bibnamefont{Keller}} \bibnamefont{and}
  \bibinfo{author}{\bibfnamefont{C.}~\bibnamefont{Bustamante}},
  \bibinfo{journal}{Biophys. J.} \textbf{\bibinfo{volume}{78}},
  \bibinfo{pages}{541} (\bibinfo{year}{2000}).

\end{thebibliography}

\end{document}